\documentclass[preprint]{aastex61}

\usepackage{longtable}

\submitjournal{\apj}

\shorttitle{Dynamic Masses of ETGs}
\shortauthors{Chen \& Hwang}

\begin{document}

\title{Dynamics of Companion Galaxies of Early-Type Galaxies}

\correspondingauthor{Chorng-Yuan Hwang}
\email{hwangcy@astro.ncu.edu.tw}

\author{Cheng-Yu Chen}
\affil{Graduate Institute of Astronomy, National Central University,
    Zhongli, Taiwan 32001}

\author{Chorng-Yuan Hwang}
\affiliation{Graduate Institute of Astronomy, National Central University,
    Zhongli, Taiwan 32001}

\begin{abstract}

We estimated the dynamical masses of 115 early-type galaxies (ETGs) by analyzing the dynamics of satellite and companion galaxies of these ETGs. 
We selected galaxies with absolute magnitudes between -22 and -25 in the $K_s$-band from the Extragalactic Distance Database (EDD). 
We also selected 216 spiral galaxies for comparison. 
We employed a simple model to simulate the observed dynamical mass from satellite galaxies at various distances. Our simulations showed that the dynamical masses derived from satellite galaxies with elliptical orbits would be smaller than those with circular orbits even they contain the same dark mass halos. Therefore, relationships between the observed $M_\mathrm{dyn}/M_\mathrm{b}$ distributions and distances would depend on orbital shapes. From the relationships between our observed $M_\mathrm{dyn}/M_\mathrm{b}$ distributions and distances, we suggest that the satellite galaxies of the ETGs have relatively more elliptical orbits than those of the spiral galaxies have and the $M_\mathrm{dyn}/M_\mathrm{b}$ of the ETGs are greater than that of the spiral galaxies.

\end{abstract}


\section{Introduction} \label{sec:intro}

The existence of dark matter phenomena has been known for more than 50 years. 
These phenomena were first detected from flat rotation curves in spiral galaxies \citep{1937ApJ....86..217Z, 1970ApJ...159..379R, 1975ApJ...201..327R, 2000MNRAS.311..441C}. 
Flat rotation curves suggest that invisible mass in the outer part is producing additional gravitational force. 
Several studies have proposed that dark matter is the dominant source of mass in spiral galaxies \citep{1973ApJ...186..467O, 1979ARA&A..17..135F, 2000MNRAS.311..441C, 2012PASJ...64...75S}.

Early-type galaxies (ETGs) have different evolution channels related to mass \citep{2016ARA&A..54..597C,2017MNRAS.468.3883P,2018MNRAS.480.1705L}.
Dwarf ETGs might form from ram-pressure stripping \citep {2011A&A...526A.114T} or gas accretion of dwarf spiral galaxies \citep{2017ApJ...840...68G,2017MNRAS.468.2850J}.
Normal ETGs ($M_* \leqslant 2 \times 10^{11} M_{\sun}$) could form from major mergers of spiral galaxies \citep{2006ApJ...650..791C,2017MNRAS.468.3883P} or star formation quench after the bulge growing up  \citep{2009ApJ...707..250M,2016ARA&A..54..597C}.
High-mass ETGs ($M_* \geqslant 2 \times 10^{11} M_{\sun}$) might form from dry mergers of low-mass ETGs \citep{2006ApJ...640..241B,2016ARA&A..54..597C,2017MNRAS.468.3883P,2018MNRAS.480.1705L} or major mergers of spiral galaxies \citep{2010MNRAS.406.2405B}.
ETGs are expected to contain a large amount of dark matter remaining from progenitor spiral galaxies.
However, several studies have revealed that visible matter is dominated in the inner regions of ETGs; for example, \citet{2014MNRAS.439.1781S} found that there are 65 elliptical galaxies showing little or no dark matter.
Moreover, several other studies also discovered that individual ETGs possess little dark matter in their inner regions \citep{2003Sci...301.1696R, 2010ApJ...719.1481V, 2011ApJ...727...96R,2015A&A...574A..93L, 2015MNRAS.446...85N, 2016MNRAS.460.3029B,2020MNRAS.491.1690J}.
Conversely, \citet{2013MNRAS.428.2407W} discovered that red galaxies have more concentrated dark halos relative to blue galaxies.
The phenomenon of little dark matter in some individual ETGs may be due to the nonexistence of dark matter \citep{2014MNRAS.439.1781S},
galaxy evolution or the spatial distribution of dark matter.
\citet{2005Natur.437..707D} simulated the merging process and concluded that the low-velocity dispersions observed in the study by \citet{2003Sci...301.1696R} were due to the radial orbits of halo stars.
Dark matter might have different distributions for different galaxies. For example, dark matter were expected to follow the Navarro-Frenk-White (NFW) model \citep{1996ApJ...462..563N} (or other distribution models), whereas baryon matter of ETGs and spiral galaxies may follow the deprojected de Vaucouleurs' model and the exponential model, respectively. These models may have different effective radii and cause variance in dark matter ratios at different radii \citep{2015ApJ...801L..20C}.
On the other hand, several studies have revealed that dark matter is dominanted in the outer region of ETGs via globular cluster kinematics \citep{2016MNRAS.458L..44F, 2016MNRAS.460.3838A,2017MNRAS.468.3949A} or satellite galaxy dynamics \citep[e.g.][]{2002ApJ...571L..85M,2011MNRAS.410..210M,2013MNRAS.428.2407W,2019MNRAS.487.3112L}. 

Satellite galaxy dynamics have been analyzed in many studies to measure the dark matter abundance on halo scales.
\citet{2002ApJ...571L..85M} analyzed 618 isolated galaxies surrounded by 1225 faint satellite galaxies in the Sloan Digital Sky Survey (SDSS) and concluded that the halo masses and central galaxy luminosities correlated linearly.
\citet{2003ApJ...593L...7B} tested the relation described by \citet{2002ApJ...571L..85M} using satellite kinematics in the Two-Degree Field Galaxy Redshift Survey (2dFGRS);
the study showed that for the spiral host, the halo mass is independent of the central luminosity.
\citet{2003ApJ...598..260P} also analyzed the satellite dynamics of isolated galaxies in the SDSS and concluded that the line-of-sight velocity dispersions of satellite galaxies decreased with the distance to the central galaxies and that the decline is consistent with the predictions of cosmological models.
\citet{2003ApJ...598..260P} also found that the relation between satellite velocity dispersions and central absolute magnitudes is very close to the Tully-Fisher relation for normal spiral galaxies.
\citet{2007ApJ...654..153C} analyzed satellite dynamics by combining data from the DEEP2 Galaxy Redshift Survey and the SDSS and found that the red central galaxies have more massive halo masses than blue central galaxies for fixed luminosity.
\citet{2011MNRAS.410..210M} also used the SDSS to analyze satellite dynamics and found that red hosts have more massive halos than blue hosts for the same luminosity, and that there is no significant difference in the average halo mass of the red and blue hosts with the same stellar mass.
\citet{2019MNRAS.487.3112L} analyzed satellite kinematics using the seventh data release (DR7) of the SDSS and observed that red central galaxies have more massive halos than blue ones of the same luminosity, and the halo mass is correlated with the average luminosity of the central galaxies.

In this study, we analyzed the dynamics of satellite galaxies of ETGs to derive the total dark matter associated with ETGs and compared these dynamics with those of spiral galaxies.
Our data selection process is described in Section 2.
We derive the dynamical masses of ETGs and compare them with those of spiral galaxies in Section 3.
Finally, we discuss the implications of mass distribution for galaxy evolution and summarize the findings of our study in Section 4.
The Hubble constant adopted in this study is $H_0 = 70~\mathrm{km}~\mathrm{s}^{-1}~\mathrm{Mpc}^{-1}.$

\section{Data} \label{sec:data}

We selected our galaxy samples from the Extragalactic Distance Database \citep[EDD;][]{2009AJ....138..323T}. 
The EDD contains tables from various sources, including the Updated Nearby Galaxy Catalog \citep{2013AJ....145..101K}, the galaxy group catalog of the Two Micron All Sky Survey \citep[2MASSGGC;][]{2015AJ....149..171T}, the third edition of the Cosmicflows database \citep{2016AJ....152...50T}, and the Arecibo Legacy Fast ALFA HI catalog \citep{2018ApJ...861...49H}. 
We obtained the $K_s$-band magnitudes and distances from EDD.
Galaxies in the local universe (less than $50~\mathrm{Mpc}$) were selected so that their small satellite galaxies could be observed (i.e. $M_{Ks}$ brighter than -21.7 at $50~\mathrm{Mpc}$ corresponding to the limiting magnitude of the 2MASSGGC $\approx 11.75$).
We used galaxies with absolute magnitudes between -22 and -25 in the $K_s$-band as our host galaxy samples.
We considered galaxies with a distance of less than $250~\mathrm{kpc}$ from a host galaxy as neighboring galaxies and selected systems that had only one neighboring galaxy to avoid groups or clusters of galaxies. 
Spiral galaxies and their companion galaxies with the same criteria were also selected for comparison.
We obtained the morphological types of these galaxies from the NASA/IPAC Extragalactic Database.
We did not select any mergers or irregular galaxies in the host galaxy and companion galaxy samples to avoid strong interacting galaxies.
The minimum projected distance between the host galaxies and companion galaxies was approximately $16~\mathrm{kpc}$.

We considered the baryonic mass $M_\mathrm{b} = L_K \times M/L_K + M_g$, where $L_k$ is the luminosity of the $K_s$-band, $M/L_K$ is the stellar mass-to-light ratio in the $K_s$-band, and $M_{g}$ is the gas mass. 
\citet{1996A&A...312..397G} used 928 spiral galaxies from eight clusters with distances between $11$ and $110~\mathrm{Mpc}$ and noted that near-infrared $M/L_K$ was almost constant. 
\citet{2003ApJS..149..289B} estimated the stellar mass functions of galaxies in the local universe using a `diet' Salpeter initial mass function (IMF) and reported that $M/L_K=0.95 \pm 0.03$ and that most of the stellar mass in the local Universe is in ETGs. 
\citet{2014AJ....148...77M} studied the color–$M/L$ relation of disk galaxies in the local universe using four population synthesis models \citep{2003ApJS..149..289B,2004MNRAS.347..691P,2009MNRAS.400.1181Z,2013MNRAS.430.2715I}, which were assumed a scaled Salpeter IMF, \citet{1998ASPC..134..483K} IMF, \citet{2003ApJ...586L.133C} IMF, and Kroupa IMF, respectively;
the four models were revised for self-consistency, and the study reported $M/L_K \approx 0.6$.
We adopted stellar $M/L_K=0.95$ and $0.6$ for our ETG and spiral galaxy samples, respectively.

The gas mass was estimated using $M_g = 1.33[M\mathrm{(HI)}+M\mathrm{(H_2)}]$, where the factor of 1.33 was due to the helium abundance \citep{2014A&A...564A..66B,2015ApJ...802...18M}.
The HI mass was obtained from the 21-cm flux using $M\mathrm{(HI)}=2.36\times10^5D^2\mathrm{flux(HI)}$ \citep{2015ApJ...802...18M}.
The $\mathrm{H_2}$ masses of some galaxies were obtained from literature \citep{1989ApJS...70..699Y, 2003PASP..115..928K, 2007PASJ...59..117K, 2008AJ....136.2782L}.
For the galaxies without available $\mathrm{H_2}$ mass, we used the relationship between $M\mathrm{(H_2)}$ and $M\mathrm{(HI)}$ described by \citet{2014A&A...564A..66B} to obtain the molecular gas mass.
For the ETGs, the atomic and molecular gas contents are much smaller than the stellar mass, and the different molecular mass estimates do not affect the final $M_\mathrm{dyn}/M_\mathrm{b}$.
For the spiral galaxies, the atomic gas content is usually less than $30\%$ of the stellar mass, and the molecular gas content is $30\%$ of the atomic gas on an average \citep{2014A&A...564A..66B}; therefore, the different molecular mass estimates can only cause a difference of a few percent in the final $M_\mathrm{dyn}/M_\mathrm{b}$ value.

We determined the dynamical mass from the relative velocity of the companion galaxy in the line of sight.
In the random distribution, the true velocity $v$ of a galaxy is related to the line-of-sight velocity $v_{\mathrm{los}}$ by the equation $v^2=3v^2_{\mathrm{los}}$.
Therefore, the dynamical mass is related to the line-of-sight velocity as $GM_\mathrm{dyn}/d = 3v_{\mathrm{los}}^2$, where $G$ is the gravitational constant, $d$ is the projected radius between the host galaxy and the neighboring galaxy, and $v_{\mathrm{los}}$ is the relative velocity of the companion galaxy on the line of sight.
The formula used for estimating dynamical mass is based on an idealized case.
The distance between a companion and host galaxy is larger than the radius of each galaxy.
The mass of the host galaxy within the satellite galaxy's orbit can thus be considered a point source at the center of the host galaxy if no dark halo surrounds the host galaxy.
However, if a dark matter halo surrounds the galaxy, our method can be used to estimate the dark matter up to the distance between a companion and host galaxy if the dark halo is spherically symmetric.
Therefore, the derived results are unaffected by the shape, rotation, or brightness profiles of the host galaxies \citep{2017MNRAS.469.2335C,2019MNRAS.488.1320N}.

The results might be affected by the inclination angles of the orbits of satellite galaxies.
However, for numerous galaxies, the viewing angles are uniformly distributed in the 3D space.
Therefore, the $M_\mathrm{dyn}/M_\mathrm{b}$ distributions measured are affected only by the real $M_\mathrm{dyn}/M_\mathrm{b}$ value.
If the observed $M_\mathrm{dyn}/M_\mathrm{b}$ distributions of ETGs are similar to the distributions of spiral galaxies, the real $M_\mathrm{dyn}/M_\mathrm{b}$ of ETGs and spiral galaxies should be similar to each other.
Dark matter constitutes approximately $90\%$ of the total mass in spiral galaxies \citep{1983Sci...220.1339R,2002ConPh..43...51K,2014RvMP...86...47C,2019MNRAS.490.5451D}.
We assumed that the upper limit of the dark matter fraction of ETGs is $90\%$;
therefore, for a galaxy containing $90\%$ dark mass, $M_\mathrm{dyn}/M_\mathrm{b}=30$ was the maximum value of the formula used to determine dynamical mass (Equation \ref{M1}).
We then only selected the galaxies that were considered to be gravitationally bounded, namely those in which $M_\mathrm{dyn}<30M_\mathrm{b}$, to be our final samples.
Our final samples comprised $115$ ETGs and $216$ spiral galaxies.
However, the chosen cutoff point might affect the final results.
We discuss the effects of using various cutoff points in Section 4.

\section{Result} \label{sec:result}

We analyzed the dynamical masses of ETGs and compared them with those of spiral galaxies.
Figure \ref{fig:morphology} displays the distributions of the dynamical-to-baryonic mass ratio $M_\mathrm{dyn}/M_\mathrm{b}$ of both ETGs and spiral galaxies.
We then used the Kolmogorov–Smirnov (K--S) test to determine whether the two $M_\mathrm{dyn}/M_\mathrm{b}$ ratios were from the same population.
The results revealed a $39\%$ chance of indistinguishable distribution of $M_\mathrm{dyn}/M_\mathrm{b}$ between the ETGs and spiral galaxy samples, which indicated a similar $M_\mathrm{dyn}/M_\mathrm{b}$ of ETGs and spiral galaxies.

Our galaxy samples had absolute magnitudes between -22 and -25.
We compared the $M_\mathrm{dyn}/M_\mathrm{b}$ distribution of galaxies between different magnitude ranges to investigate the relation between luminosity and dynamical mass.
Figure \ref{fig:luminosity} displays the $M_\mathrm{dyn}/M_\mathrm{b}$ distributions of galaxies with different absolute magnitudes.
The K--S test results of these distributions have $p$-value $>0.05$ and indicate that the $M_\mathrm{dyn}/M_\mathrm{b}$ distributions of galaxies with different absolute magnitudes were indistinguishable.

We analyzed the $M_\mathrm{dyn}/M_\mathrm{b}$ values estimated at various distances to investigate the spatial distributions of dark matter.
Figure \ref{fig:distance} presents the $M_\mathrm{dyn}/M_\mathrm{b}$ distributions measured from the companion galaxies at different distances from the host galaxies for both morphologies.
Table \ref{tab:KStest} lists the distribution results of the K--S test, which indicated that the $M_\mathrm{dyn}/M_\mathrm{b}$ distributions of closer and farther companion galaxies of the ETGs were not substantially different.
However, the farther companion galaxies of spiral galaxies had larger $M_\mathrm{dyn}/M_\mathrm{b}$ values than the closer satellite galaxies of spiral galaxies. 
We also tested the correlation with Kendall's tau,which is a non-parametric measure of the correlation between two ranked parameters.
The null hypothesis of Kendall's rank test is that the two parameters are independent.
The $\tau$ and $p$-value for the ETGs are 0.013 and 0.84, respectively, whereas those for the spiral galaxies are 0.23 and $6.3\times 10^{-7}$, respectively.
These results also indicate that $M_\mathrm{dyn}/M_\mathrm{b}$ is not correlated with $d$ for the ETGs but is correlated for the spiral galaxies.

\section{Discussion} \label{sec:discu}

\begin{figure}
\plotone{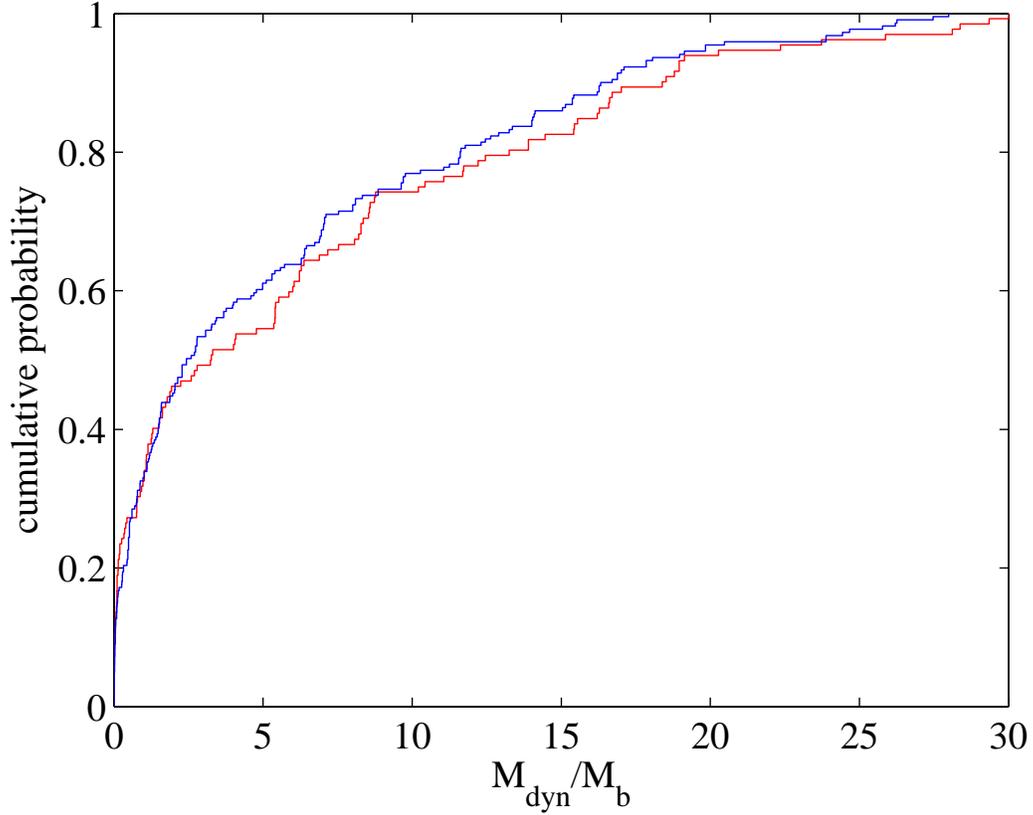}
\caption{Probability distributions of $M_\mathrm{dyn}/M_\mathrm{b}$ of ETGs and spiral galaxies. The red and blue lines represent the distributions of the ETGs and spiral galaxies, respectively.\label{fig:morphology}}
\end{figure}

We analyzed $M_\mathrm{dyn}/M_\mathrm{b}$ values from the dynamics of the early-type and spiral galaxy companions.
Figure \ref{fig:morphology} revealed that the $M_\mathrm{dyn}/M_\mathrm{b}$ distributions of ETGs and spiral galaxies were similar; the K--S test indicates that the dark matter fractions in the outskirts of  the ETGs were similar to those of the spiral galaxies.
We also obtained the same conclusions from the 2-sample Anderson-Darling (A--D) test. 
The test statistic for the A--D test is 0.65, and the $p$-value is greater than 0.1.
These results also indicate that the $M_\mathrm{dyn}/M_\mathrm{b}$ distributions of the ETGs and the spiral galaxies are indistinguishable.

\begin{figure}
\plottwo{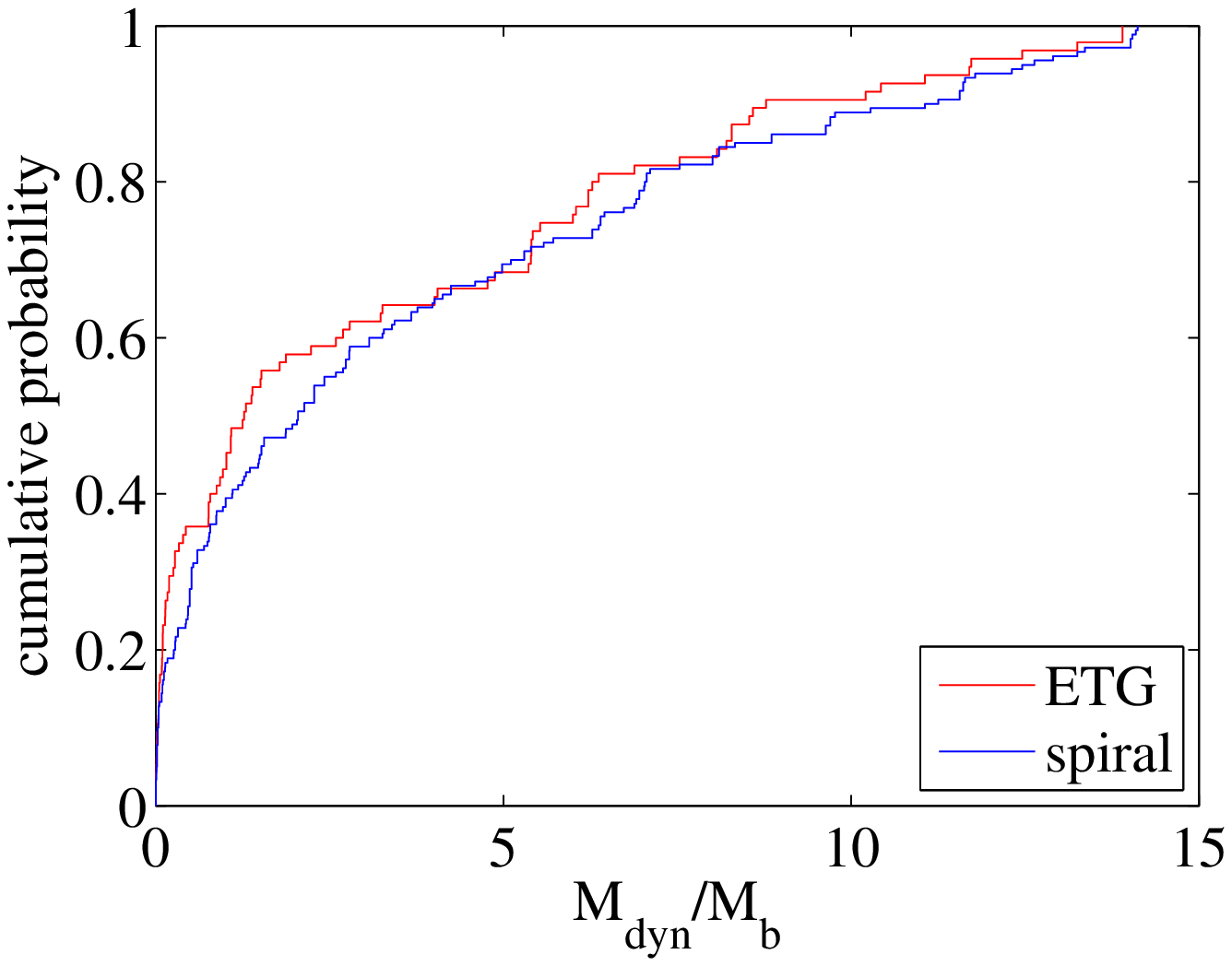}{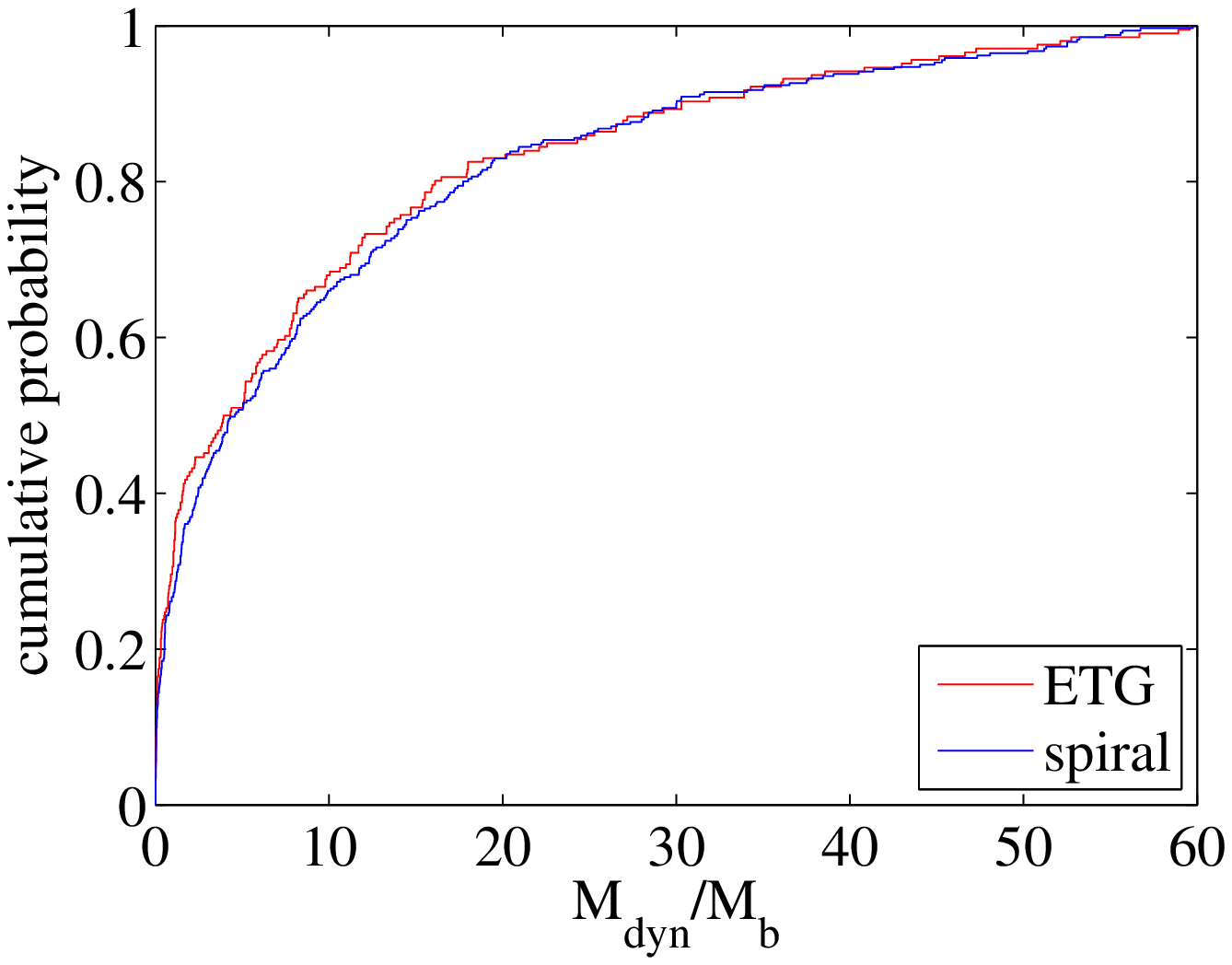}
\caption{Probability distributions of $M_\mathrm{dyn}/M_\mathrm{b}$ of the ETGs and spiral galaxies using $80\%$ and $95\%$ dark matter fraction cutoff.
Left: $80\%$ dark matter fraction cutoff.
Right: $95\%$ dark matter fraction cutoff.
The red and blue lines represent the $M_\mathrm{dyn}/M_\mathrm{b}$ distributions of ETGs and spiral galaxies, respectively.\label{fig:cutoff}}
\end{figure}

We investigate the possible influence of the dark matter fraction cutoff on our results.
We obtained the $M_\mathrm{dyn}/M_\mathrm{b}$ distributions by applying $80\%$ and $95\%$ dark matter fraction cutoffs, which correspond to $M_\mathrm{dyn}/M_\mathrm{b}$ distribution cutoffs of 15 and 60, respectively.
Figure \ref{fig:cutoff} illustrates the $M_\mathrm{dyn}/M_\mathrm{b}$ distributions obtained using various cutoff values.
We then checked the similarity using the K-S and A-D tests.
The $p$-values of the K--S test are 0.25 and 0.33 for $80\%$ and $95\%$ cutoffs, respectively, and the test statistics of the A--D test are 1.04 and 0.64 for $80\%$ and $95\%$ cutoff, respectively; both values correspond to $p$-values greater than 0.1.
These results suggest that the $M_\mathrm{dyn}/M_\mathrm{b}$ distributions of the ETGs and the spiral galaxies are similarly independent of the cutoffs.

\begin{figure}
\plottwo{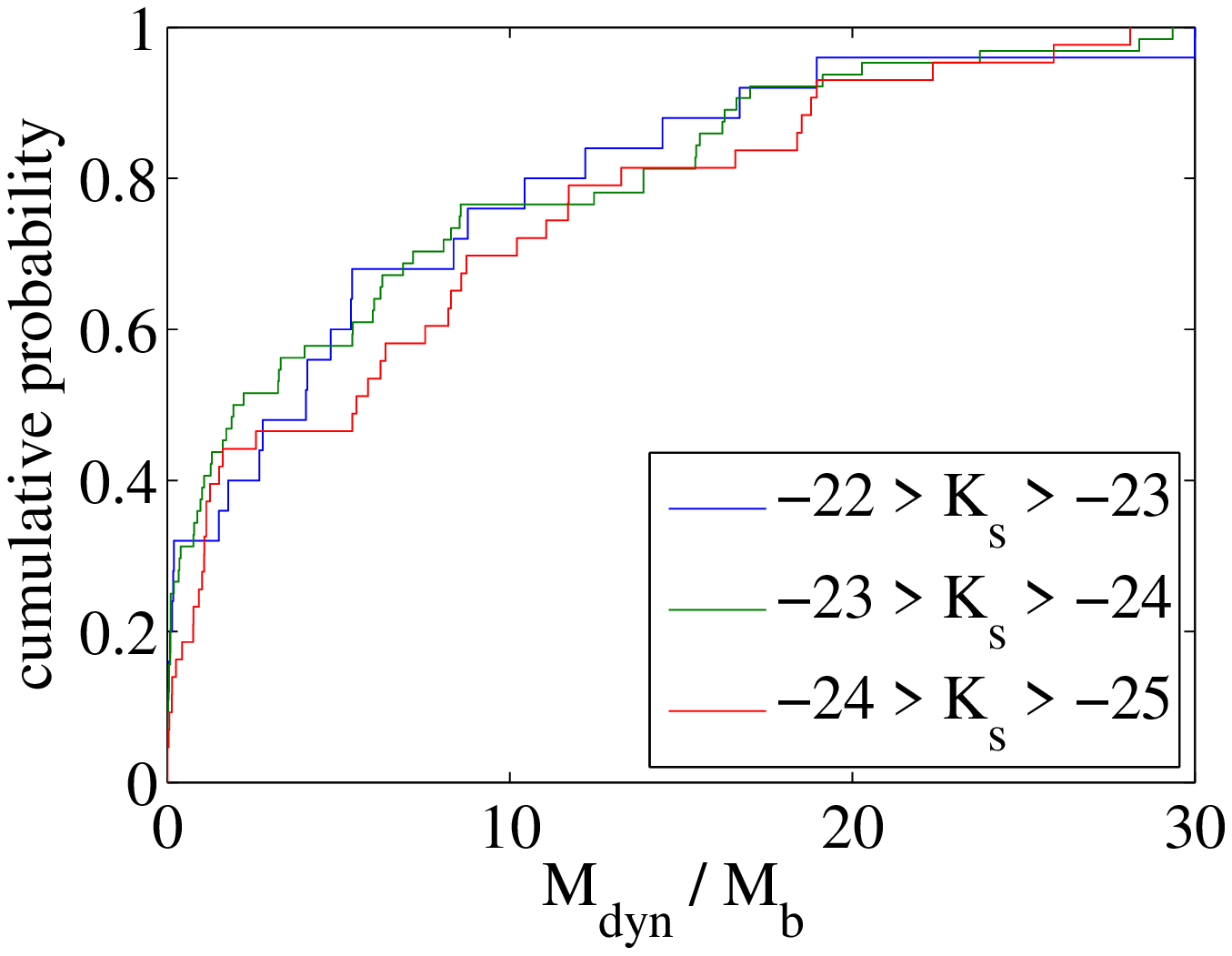}{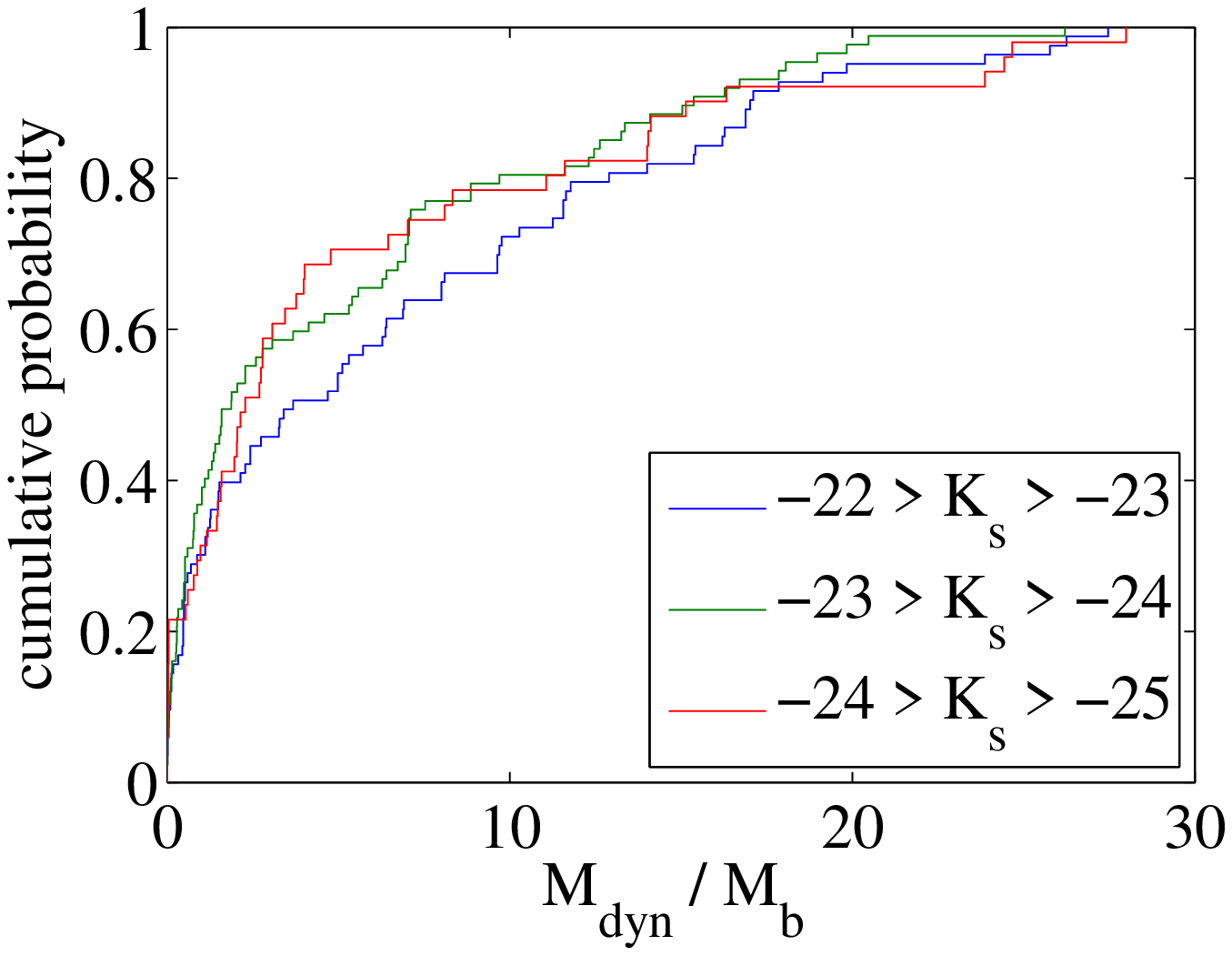}
\caption{Probability distributions of $M_\mathrm{dyn}/M_\mathrm{b}$ of the ETGs and spiral galaxies of different galaxy luminosities.
Left: $M_\mathrm{dyn}/M_\mathrm{b}$ of ETGs.
Right: $M_\mathrm{dyn}/M_\mathrm{b}$ of spiral galaxies.
The blue, green, and red lines represent the $M_\mathrm{dyn}/M_\mathrm{b}$ distributions of the galaxies that had absolute magnitudes from -22 to -23, -23 to -24, and -24 to -25, respectively.\label{fig:luminosity}}
\end{figure}

Figure \ref{fig:luminosity} shows that the $M_\mathrm{dyn}/M_\mathrm{b}$ distributions are not related to luminosities.
The brighter galaxies contain larger baryonic masses; therefore, the similar $M_\mathrm{dyn}/M_\mathrm{b}$ ratios indicate that the dark halos of the brighter galaxies are more massive.
This result is in agreement with the positive correlation between the dark halo mass and the central galaxy luminosity, as described in previous works \citep{2002ApJ...571L..85M,2003ApJ...598..260P,2019MNRAS.487.3112L}.

\begin{figure}
\plottwo{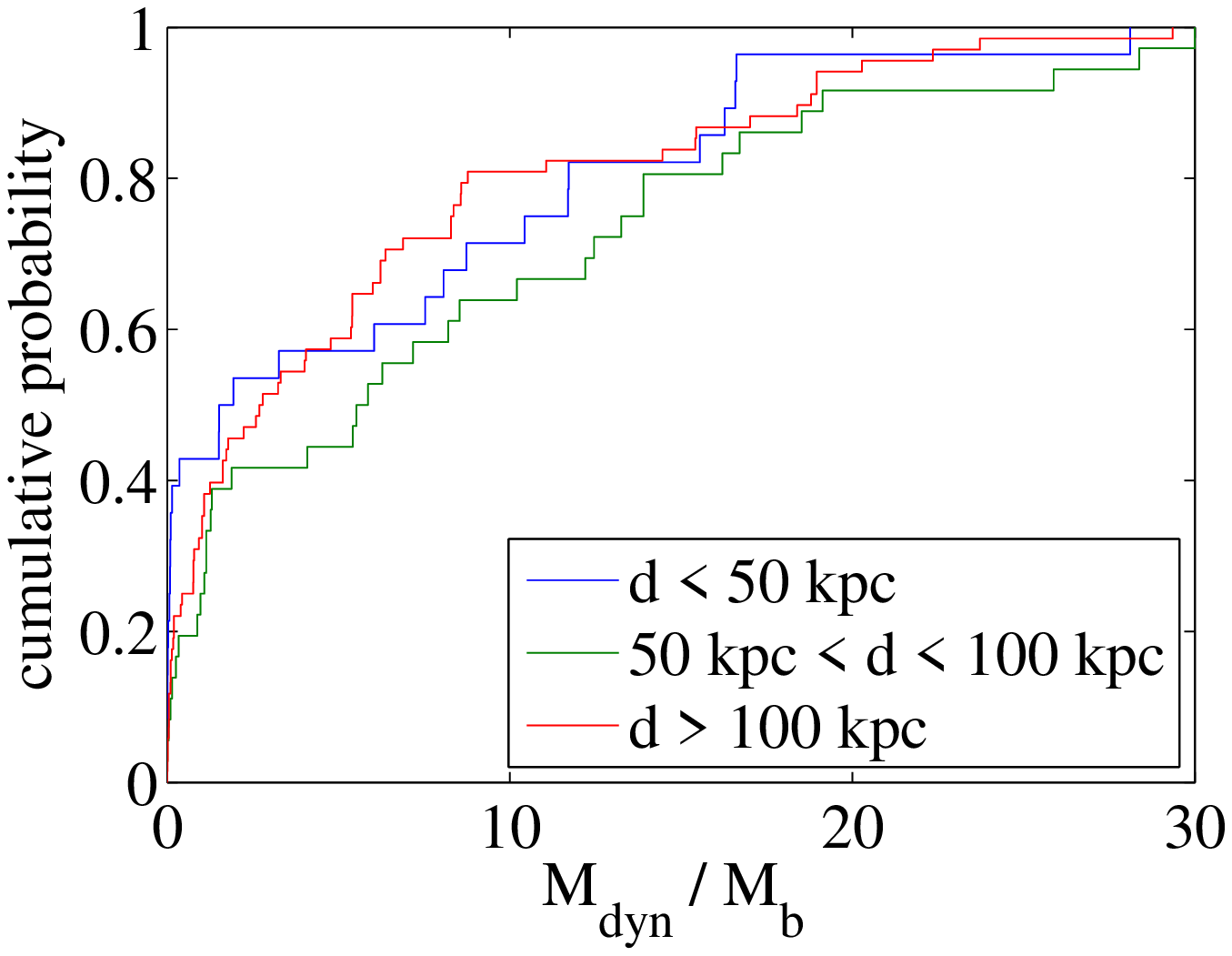}{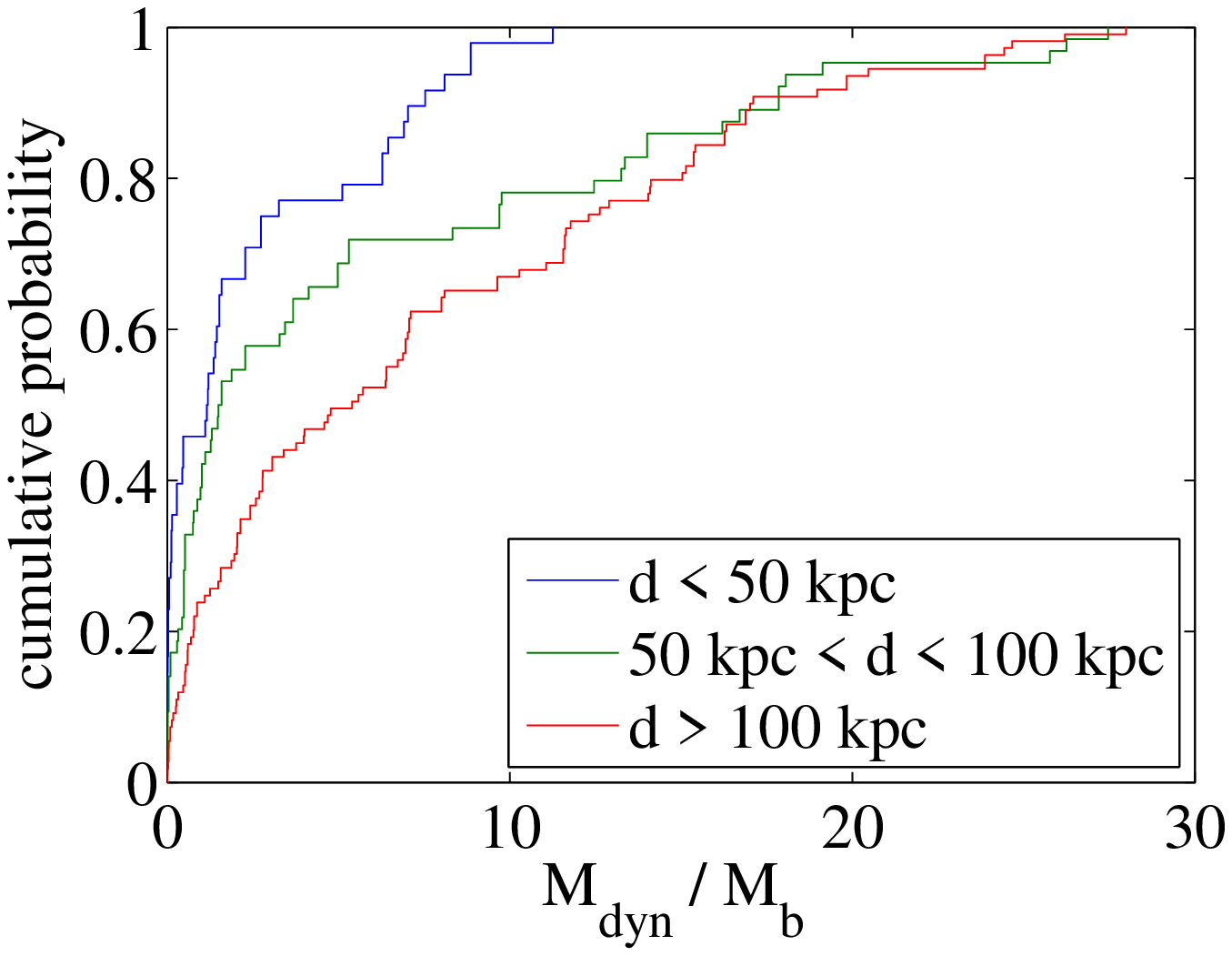}
\caption{Probability distributions of $M_\mathrm{dyn}/M_\mathrm{b}$ of the ETGs and spiral galaxies of different observed satellite-to-host galaxy distances.
Left: $M_\mathrm{dyn}/M_\mathrm{b}$ of ETGs.
Right: $M_\mathrm{dyn}/M_\mathrm{b}$ of spiral galaxies.
The blue, green, and red lines represent the $M_\mathrm{dyn}/M_\mathrm{b}$ distributions for distances less than $50~\mathrm{kpc}$, between $50~\mathrm{kpc}$ and $100~\mathrm{kpc}$, and greater than $100~\mathrm{kpc}$, respectively.\label{fig:distance}}
\end{figure}

\begin{figure*}
\gridline{\fig{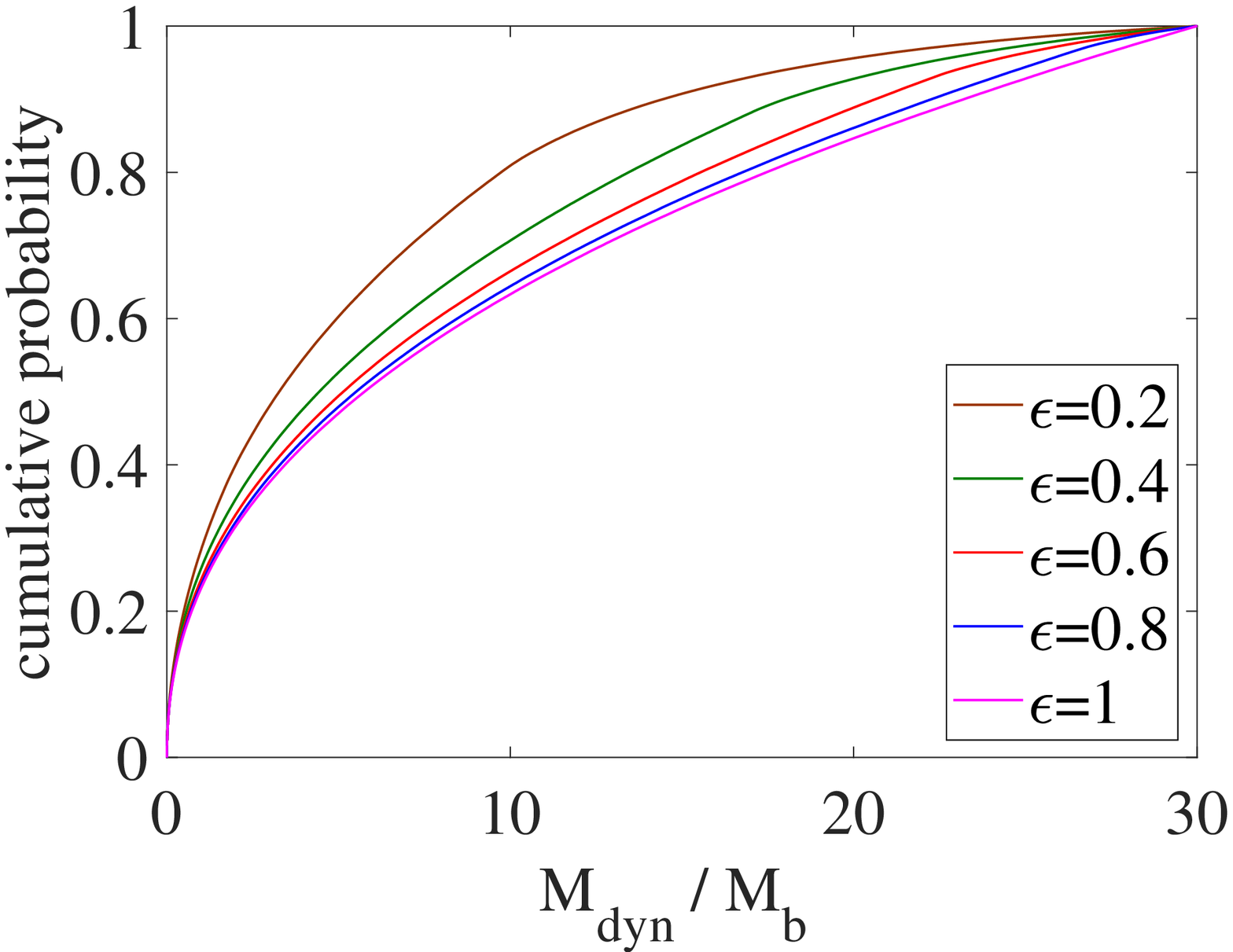}{0.3\textwidth}{(a)}
          \fig{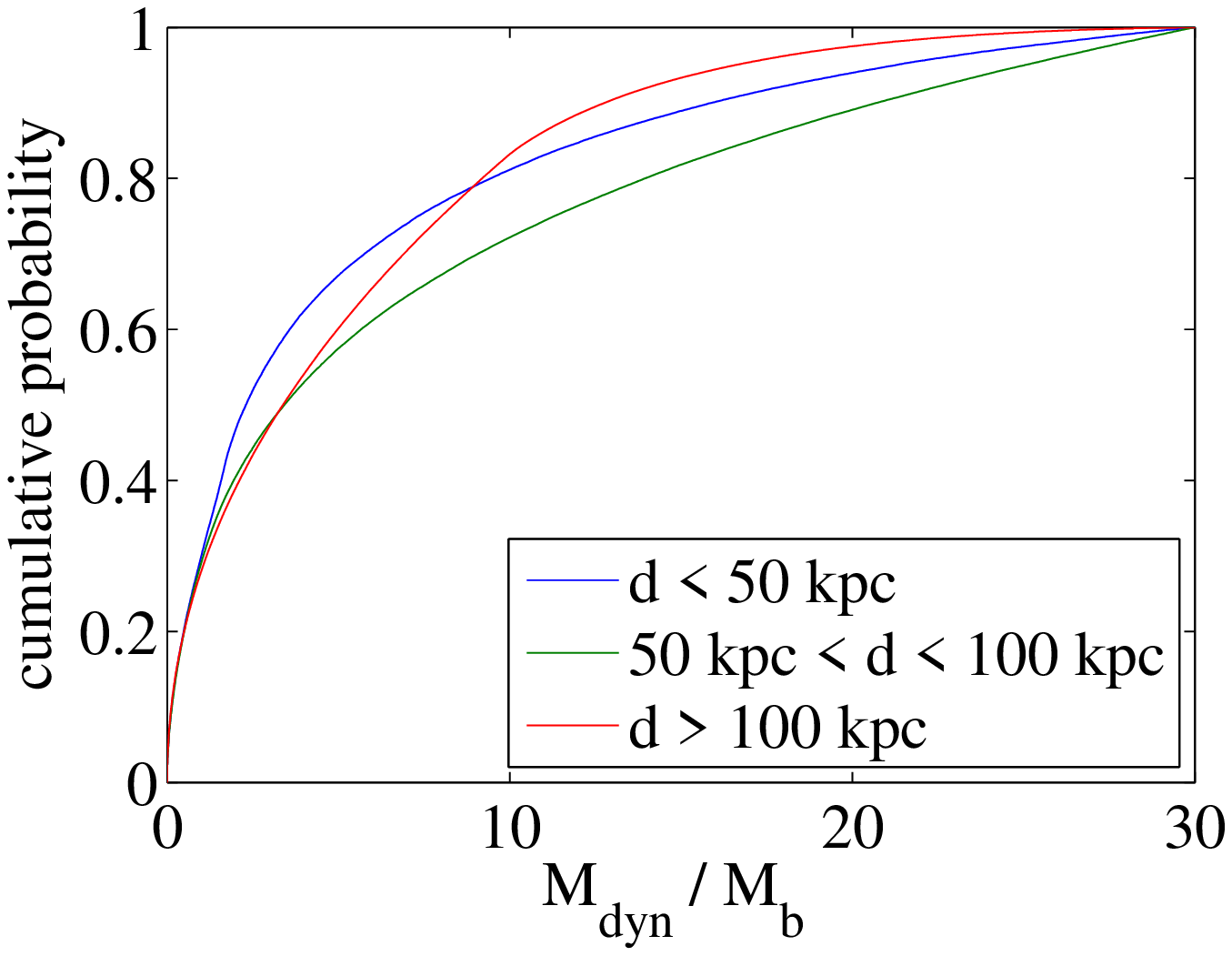}{0.3\textwidth}{(b)}
          \fig{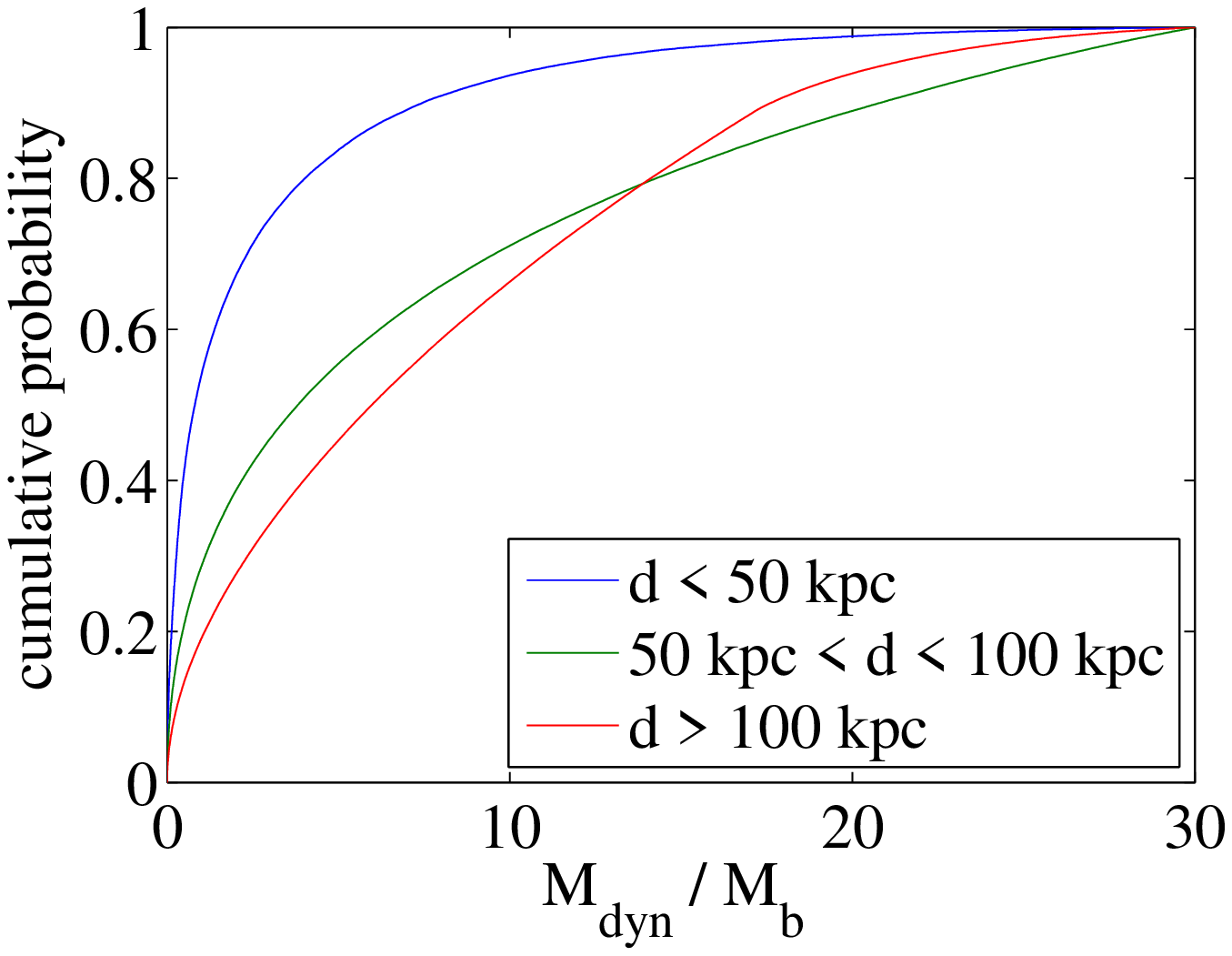}{0.3\textwidth}{(c)}}
\gridline{\fig{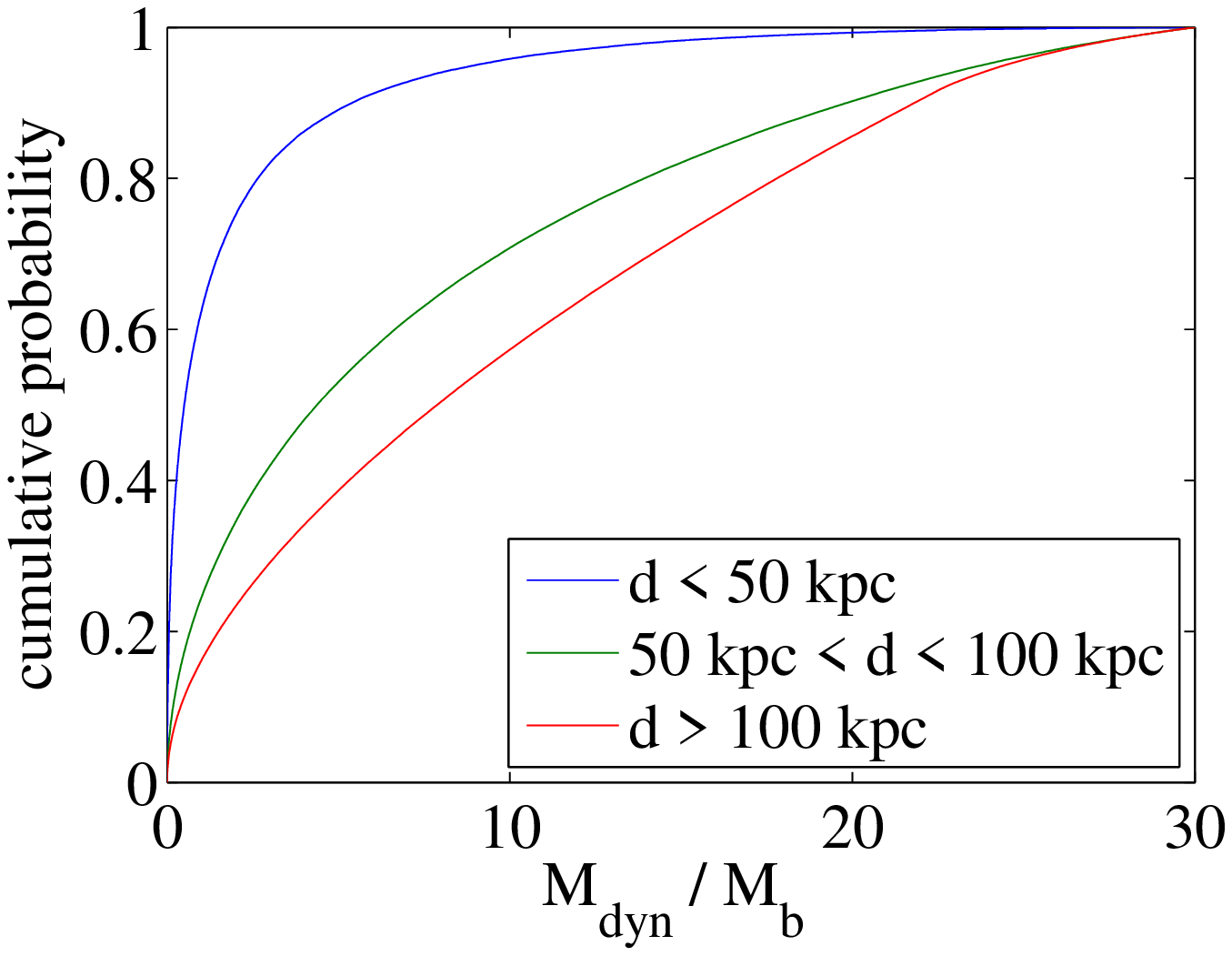}{0.3\textwidth}{(d)}
          \fig{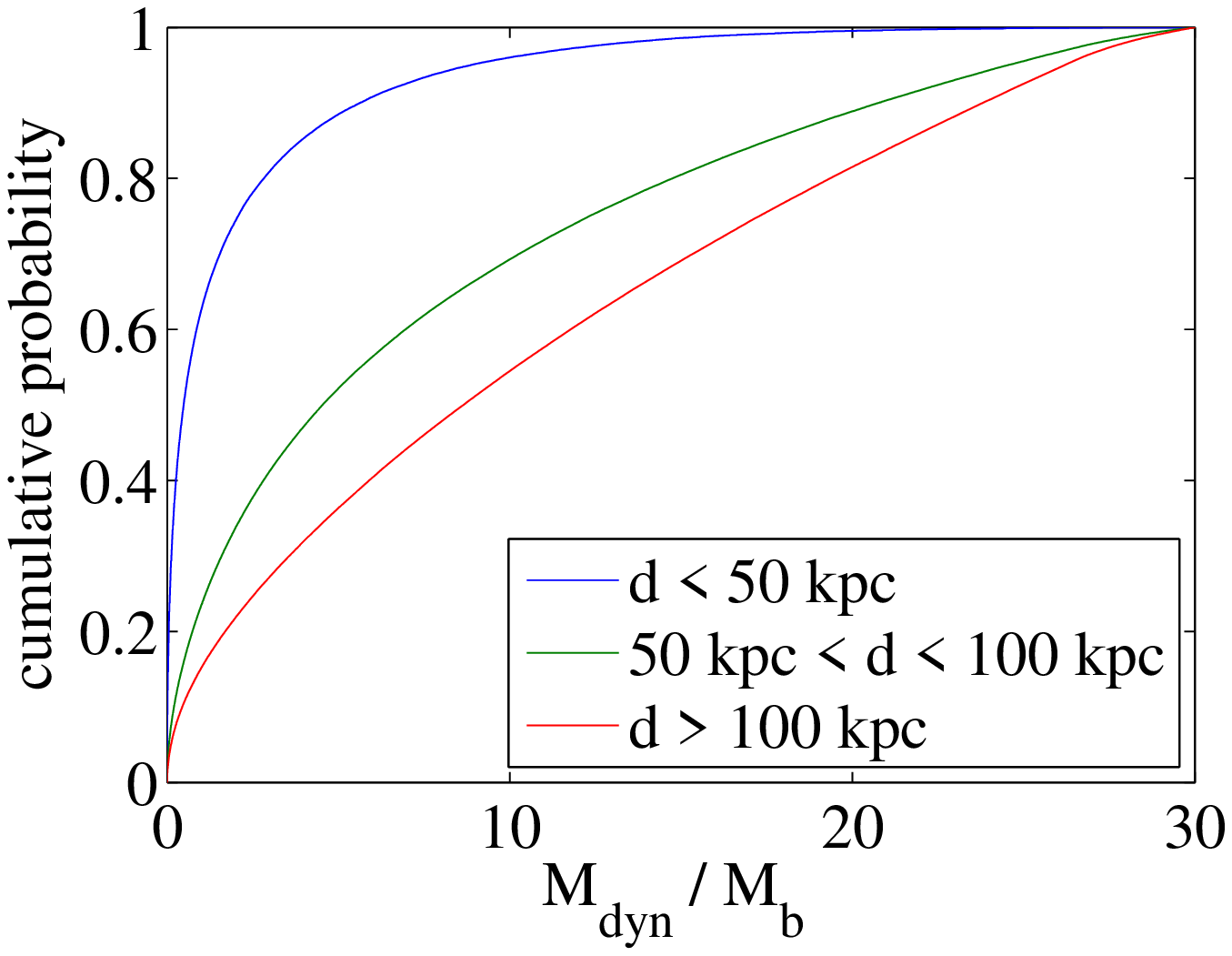}{0.3\textwidth}{(e)}
          \fig{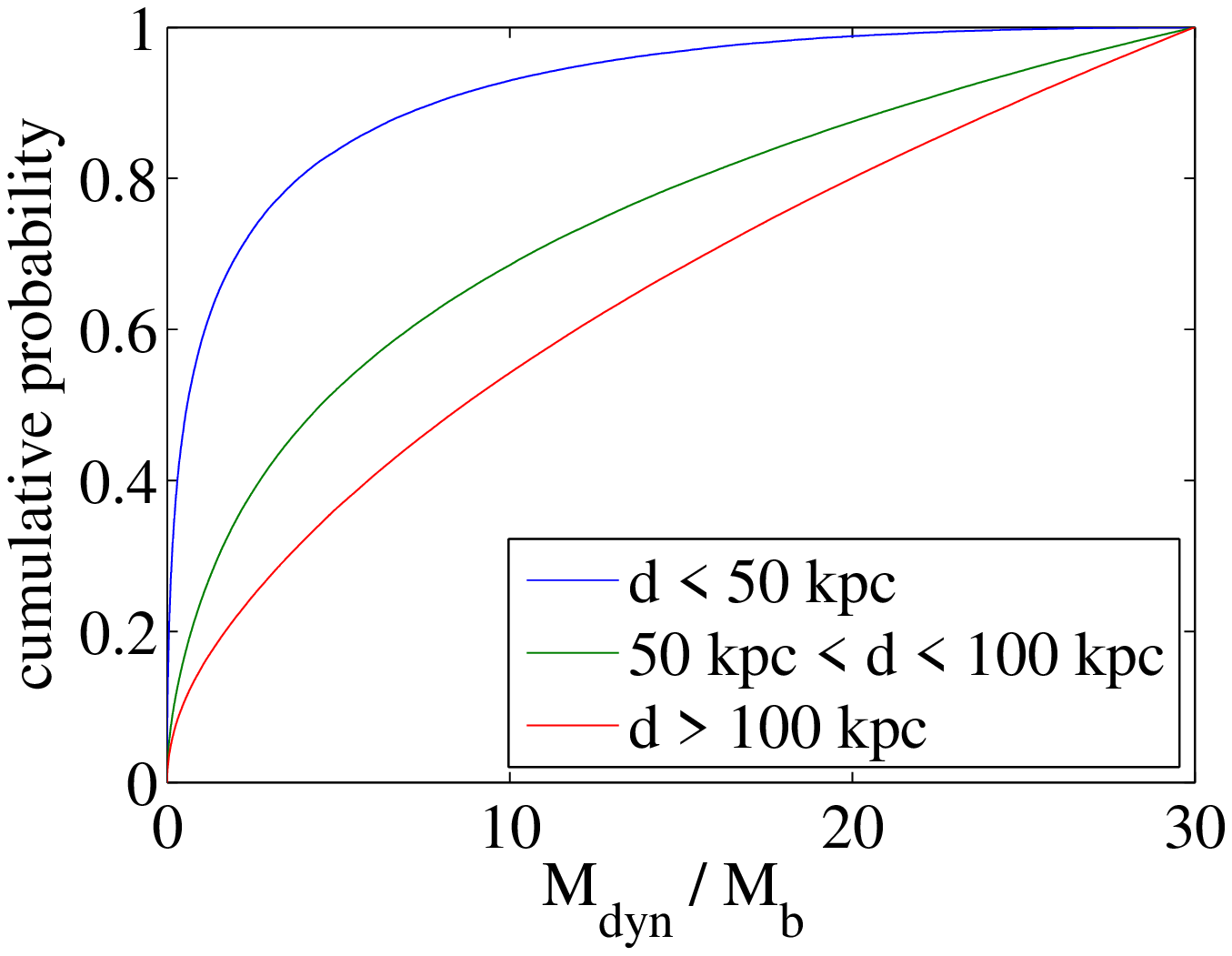}{0.3\textwidth}{(f)}}
\caption{Probability distributions of simulated $M_\mathrm{dyn}/M_\mathrm{b}$ values for different elliptical orbits. (a) Brown, green, red, blue, and magenta lines represent the results for the orbits with pericenter to apocenter length ratios ($\epsilon$) of 0.2, 0.4, 0.6, 0.8, and 1, respectively. (b–f) The $M_\mathrm{dyn}/M_\mathrm{b}$ distributions of different observed satellite-to-host galaxy distances for 
orbit with $\epsilon$ values of (b) 0.2, (c) 0.4, (d) 0.6, (e) 0.8 and (f) 1, respectively.
Blue, green, and red lines represent the results for distances less than $50~\mathrm{kpc}$, between $50~\mathrm{kpc}$ and $100~\mathrm{kpc}$, and greater than $100~\mathrm{kpc}$, respectively.
\label{fig:sim}}
\end{figure*}

\begin{figure}
\plotone{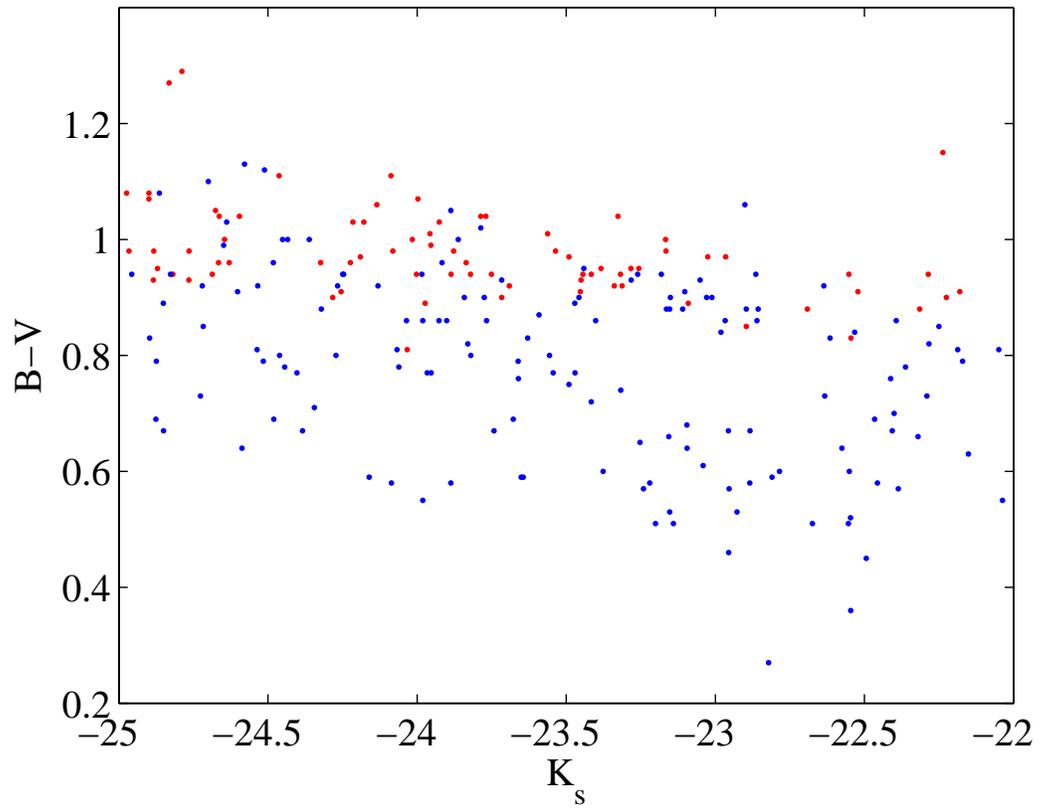}
\caption{
Color-Magnitude diagram of the host galaxies. The red and blue dots represent the distributions of the ETGs and spiral galaxies, respectively.
\label{fig:colormag}}
\end{figure}

Figure \ref{fig:distance} reveals that the $M_\mathrm{dyn}/M_\mathrm{b}$ values of ETGs were not related to the distances between the companion and host galaxies, whereas the $M_\mathrm{dyn}/M_\mathrm{b}$ values of the spiral galaxies were.
For the ETG samples, the $M_\mathrm{dyn}/M_\mathrm{b}$ values measured from the closer and farther companions were indistinguishable.
By contrast, the farther companion galaxies of spiral galaxies demonstrated larger $M_\mathrm{dyn}/M_\mathrm{b}$ values of the host galaxies than did the closer companions.
These results may indicate different dark matter density profiles for the ETGs and spiral galaxies; in this situation, the dark matter of the ETGs galaxies would be more concentrated in the inner regions, whereas that of the spiral galaxies would be expanded to the outer regions.
These results are consistent with those of \citet{2013MNRAS.428.2407W}, who discovered that the dark matter halos of red host galaxies are significantly more concentrated than those of blue hosts of the same stellar mass are.
Another possibility is that the orbits of the companion galaxies of the ETGs and spiral galaxies may have different degrees of ellipticity.
We employed a simple model to simulate the observed dynamical masses from the satellite galaxies at different elliptical orbits.
We assumed that all masses were distributed within $50~\mathrm{kpc}$, and the pericenter to apocenter length ratios $\epsilon$ of the orbits were 0.2, 0.4, 0.6, 0.8, and 1.
The details of our simulations are described in Appendix A.
The simulated results are presented in Figure \ref{fig:sim}.
The satellite galaxies with circular or nearly circular orbits exhibited considerable dynamics–distance relations; farther satellites exhibited larger $M_\mathrm{dyn}/M_\mathrm{b}$ values of the host galaxies than closer satellites.
However, the satellite galaxies with more elliptical orbits did not exhibit a dynamics–distance relationship.
These results indicated that the orbits of the companion galaxies of the ETG samples may be more elongated than those of the spiral galaxies.
\citet{2013MNRAS.428.2407W} also found that satellite orbits around red galaxies are radial anisotropic and that satellite orbits around blue galaxies are consistent with isotropic models.
The reason behind this elongation can be understood by considering the merger of a system with two host galaxies and a satellite galaxy.
When the two host galaxies merge, most of the angular moment of the system will be passed to the satellite galaxy.
The satellite galaxy will gain angular momentum from the progenitor galaxies during the merging process and will follow an elongated orbit surrounding the merged ETG.
This process is dynamically similar to the encounters between binaries and single stars; during the encounter, the two more massive stars are tightly bound, and the third one is ejected \citep{1975AJ.....80..809H}.
Similar process was reported by \citet{2005Natur.437..707D}, who simulated disk-galaxy mergers and found that the orbits of the outer stars of resulting galaxies were elongated.
Since the $M_\mathrm{dyn}/M_\mathrm{b}$ values measured from elongated orbits are smaller than those from circular orbits with the same real $M_\mathrm{dyn}/M_\mathrm{b}$, the similar $M_\mathrm{dyn}/M_\mathrm{b}$ distributions from observation data indicate that the ETGs have larger real $M_\mathrm{dyn}/M_\mathrm{b}$ than the spirals.
These result are consistent to some previous studies, which suggested that red host galaxies had more massive halos than blue ones with the same optical magnitudes \citep{2007ApJ...654..153C,2011MNRAS.410..210M,2019MNRAS.487.3112L}.
Although our host galaxies were classified based on morphologies, our ETG sample is redder than the spiral one as shown in Figure \ref{fig:colormag}.

Our model mainly considers the influence of the orbital anisotropies of satellite galaxies.
The anisotropic effects originate primarily from the viewing angles and orbital shapes.
We measured more than 100 galaxies; the probability that we observed all of these galaxies at certain viewing angles is negligible.
Therefore, the influence of viewing angles on the orbits of the satellite galaxies is minimized.
On the other hand, we would underestimate the dynamical mass of the host galaxies by considering satellite galaxies with elongated elliptical orbits.
Figure \ref{fig:sim} also illustrates the $M_\mathrm{dyn}/M_\mathrm{b}$ distribution measured from satellite galaxies with different orbital shapes and shows that the $M_\mathrm{dyn}/M_\mathrm{b}$ distribution is skewed toward lower $M_\mathrm{dyn}/M_\mathrm{b}$ for more elongated orbits.

\begin{figure}
\gridline{\fig{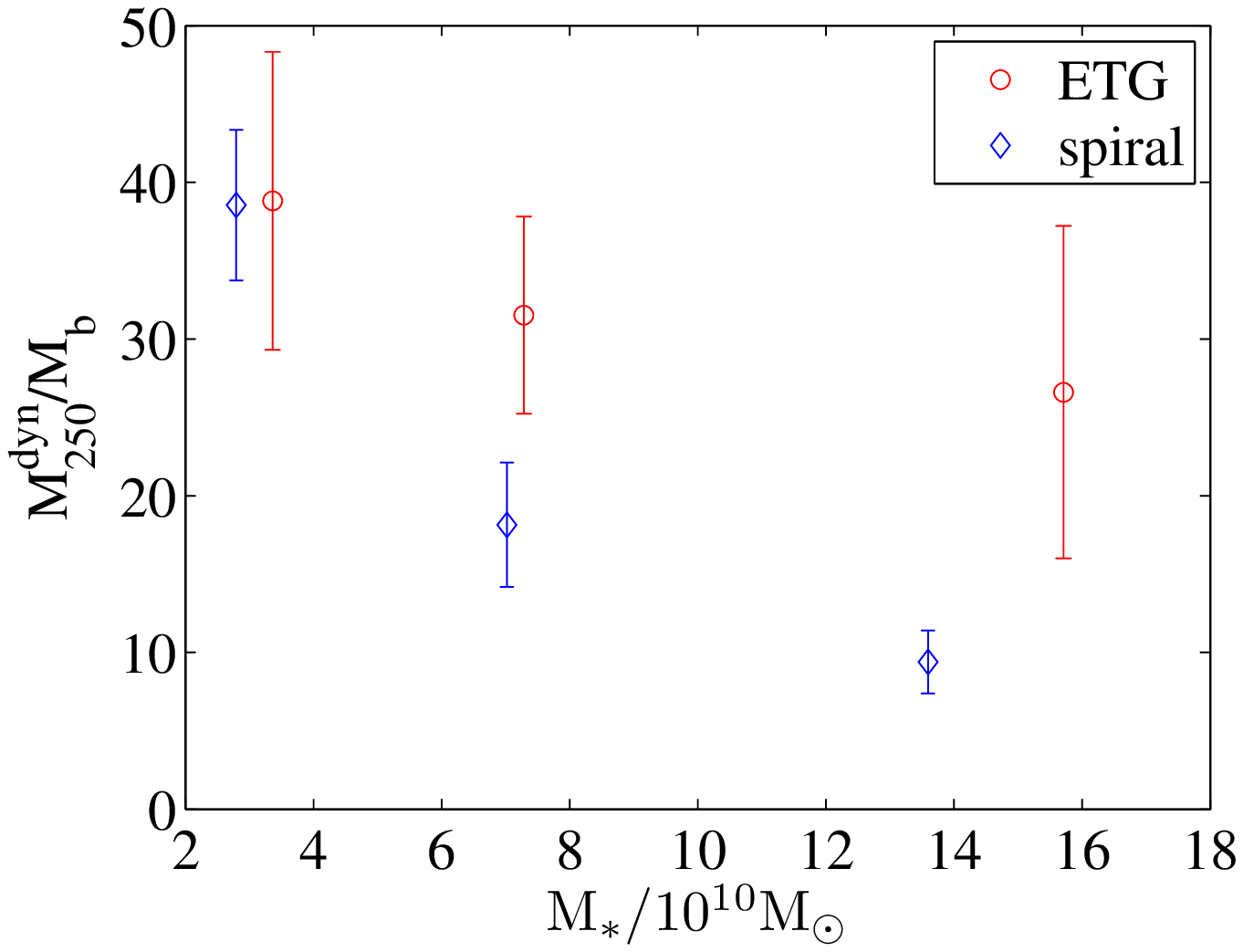}{0.5\textwidth}{(a)}
          \fig{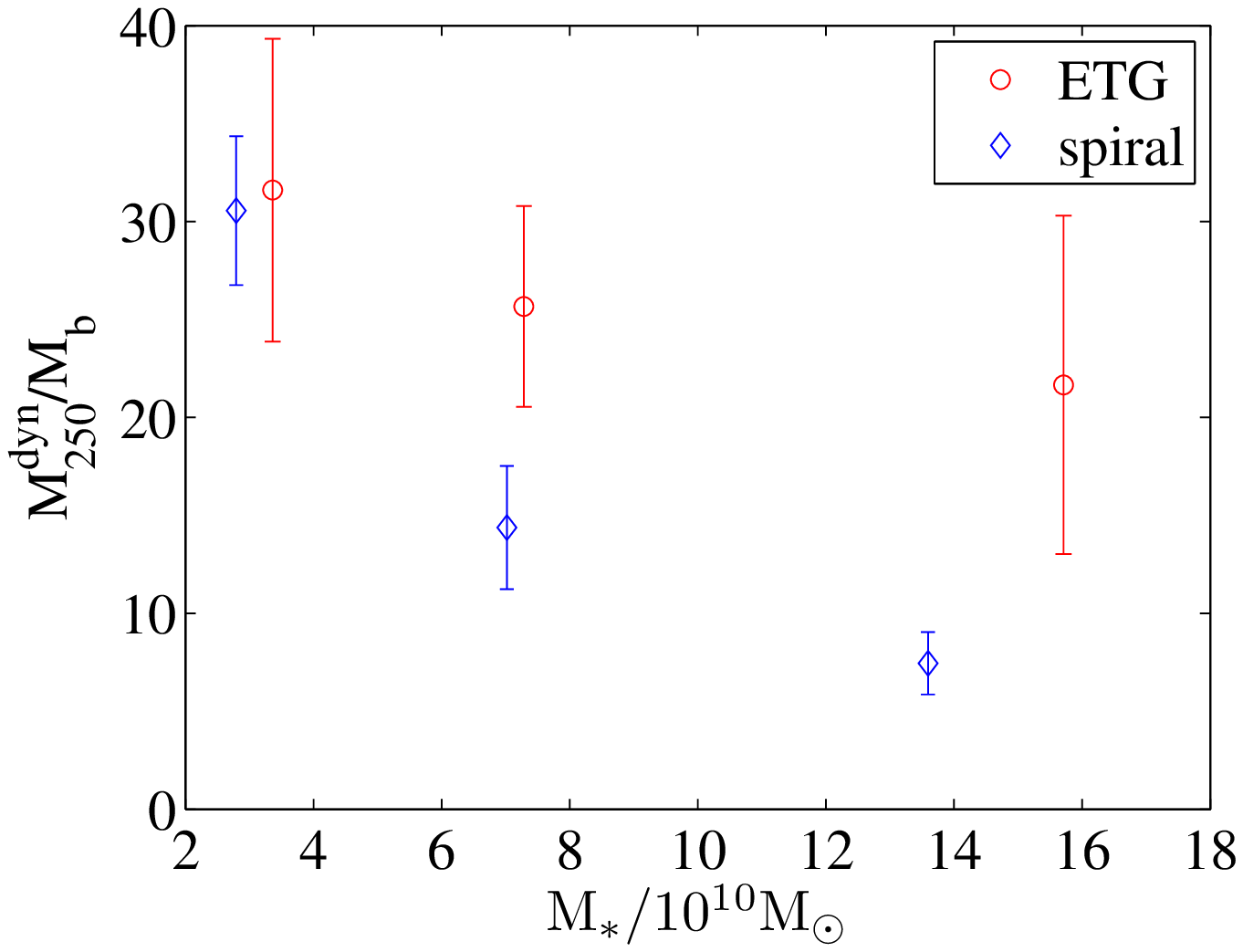}{0.5\textwidth}{(b)}}
\gridline{\fig{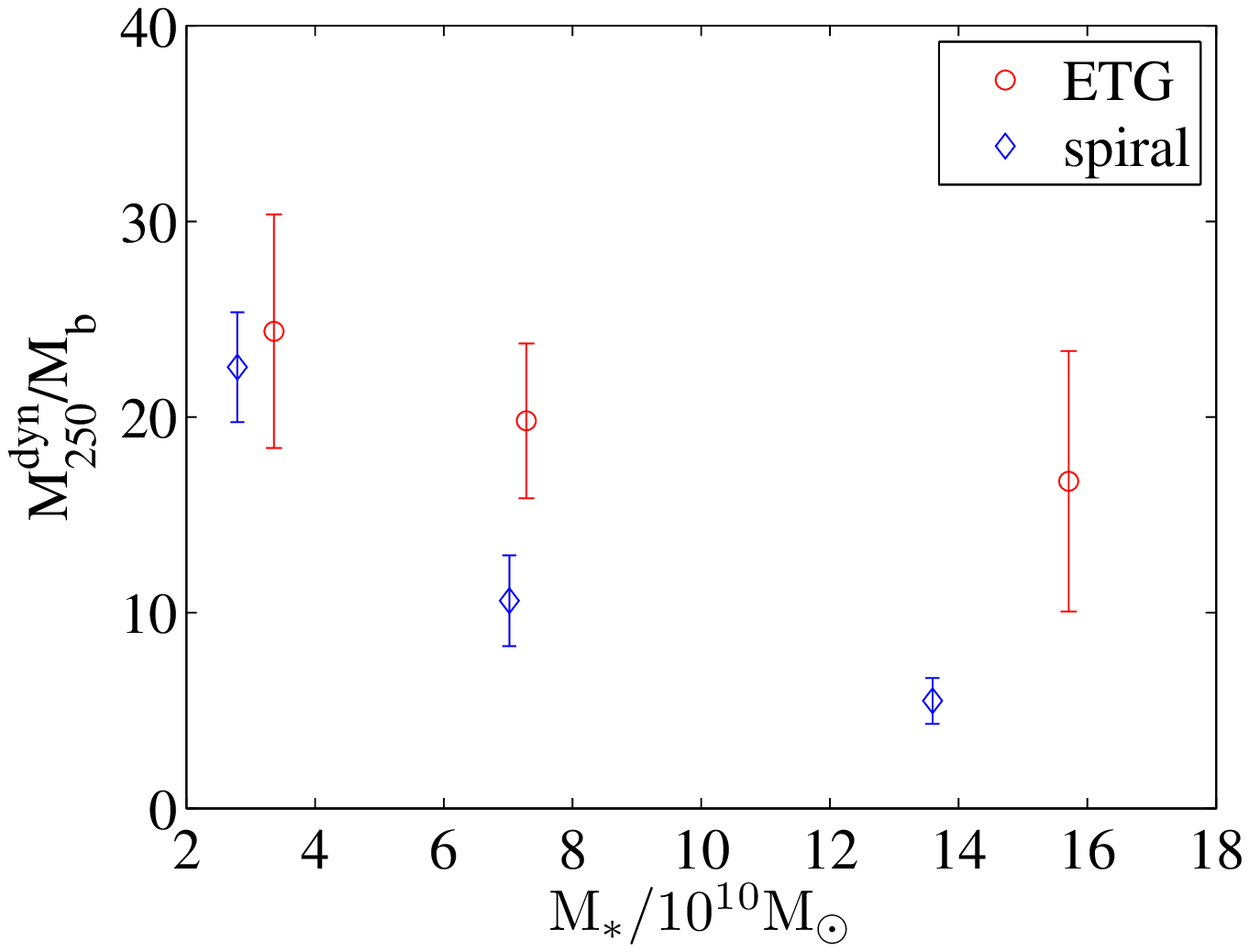}{0.5\textwidth}{(c)}
          \fig{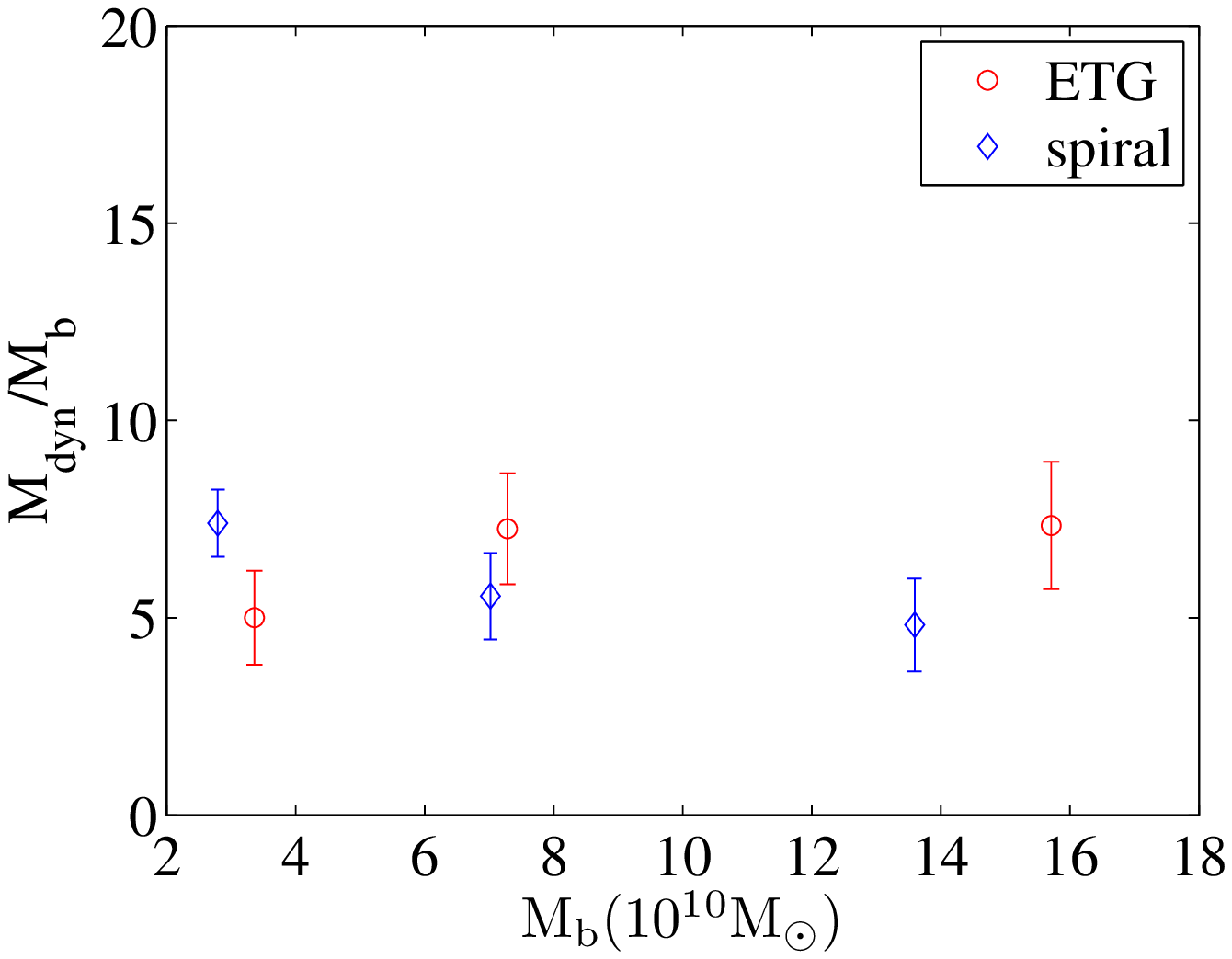}{0.5\textwidth}{(d)}}
\caption{
$M^\mathrm{dyn}_\mathrm{250}/M_\mathrm{b}$ as a function of baryon mass with various anisotropy values (a) $\beta=0$ (b) $\beta=0.25$, and (c) $\beta=0.5$.
The results derived from our statistical method are shown in (d) for comparison.
The galaxy samples were divided into three baryonic mass bins: $M_b<5\times 10^{10} \mathrm{M_\odot}$, $5\times 10^{10} \mathrm{M_\odot}<M_b<10\times 10^{10} \mathrm{M_\odot}$, and $M_b>10\times 10^{10} \mathrm{M_\odot}$.
\label{fig:MdvsMb}}
\end{figure}

We also used the method of \citet{2002ApJ...571L..85M} to derive the
dynamical mass within $250~\mathrm{kpc}$ of the host galaxies, $M^\mathrm{dyn}_\mathrm{250}$, 
for comparison to verify consistency between our method and others.
We used the spherical Jeans equation to determine the dynamical mass \citep{2008gady.book.....B},
\begin{equation}
\frac{GM(r)}{r} = -\overline{v_r^2}[\frac{d\log n(r)}{d\log r} + \frac{d\log \overline{v_r^2}}{d\log r} + 2\beta],
\label{Jeans}
\end{equation}
where $r$ is the $250~\mathrm{kpc}$ radius to which we integrate the mass, $\overline{v_r^2}$ is the radial mean square velocity, $n(r)$ is the number density of satellites as a function of radius, and $\beta$ is the velocity anisotropy $\beta = 1 - \overline{v_t^2}/\overline{v_r^2}$.
We assumed that the radial mean squared velocity $\overline{v_r^2}$ is equal to the line-of-sight mean squared velocity $\overline{v_\mathrm{los}^2}$ as \citet{2002ApJ...571L..85M} did,
who found that $\overline{v_\mathrm{los}^2}=1.03\pm0.03~\overline{v_r^2}$.
We then determined the number densities and mean square velocities for various distances between the companion and host galaxies.
Subsequently, we fitted the data to the linear relation between log number density and log distance and to the linear relation between log mean square velocity in apertures from 50 to $250~\mathrm{kpc}$ and log distance.  
The value of $\frac{d\log n(r)}{d\log r}$ is the slope of the fitted line of the density profile--distance relation, and the value of $\frac{d\log \overline{v_r^2}}{d\log r}$ is the slope of the velocity profile--distance relation.
We found that the values of $\frac{d\log n(r)}{d\log r}$ are -2.2 and -2.3 for the ETGs and the spiral galaxies, respectively, and those of $\frac{d\log \overline{v_r^2}}{d\log r}$ are -0.49 and -0.11 for the ETGs and the spiral galaxies, respectively.
The anisotropy values are usually between 0 and 0.5 \citep{2003ApJ...598..260P,2013MNRAS.428.2407W,2019MNRAS.482.4824L}; therefore, we use the anisotropy values $\beta =0$, 0.25, and 0.5.
These results are shown in Figure \ref{fig:MdvsMb}.
The $M^\mathrm{dyn}_\mathrm{250}/M_\mathrm{b}$ values of the ETGs were greater than those of the spiral galaxies.
These results ostensibly differed from the results presented in Figure \ref{fig:morphology}.
However, our simulations indicated that, on average, satellites of ETGs follow more elliptical orbits than those of spiral galaxies do.
This suggests that our method potentially underestimated the dynamical masses of the ETGs.
Therefore, ETGs, on average, have greater dynamical-to-baryonic mass ratios than spiral galaxies have.
Thus, the results derived from our method are consistent with those obtained from applying the method of \citet{2002ApJ...571L..85M}.
We note that the $M^\mathrm{dyn}_\mathrm{250}/M_\mathrm{b}$ values obtained from Equation \ref{Jeans} are much larger than the $M_\mathrm{dyn}/M_\mathrm{b}$ values obtained using the method described in Section 2.
The main reason for this is that in Equation \ref{Jeans}, the dynamical mass is estimated for the total halo mass within $250~\mathrm{kpc}$, but our method provides an estimate of the halo mass only within the projected distances for individual galaxies.
The projected distances of individual galaxies are usually much smaller than $250~\mathrm{kpc}$; smaller $M_\mathrm{dyn}/M_\mathrm{b}$ values are thus derived.
Furthermore, our data differed from the data of \citet{2002ApJ...571L..85M}, who have selected galaxy sample up to $500~h^{-1}~\mathrm{kpc}$, and we thus obtained a different mean square velocity profile and different $\frac{d\log \overline{v_r^2}}{d\log r}$ value.
However, this difference in $\frac{d\log \overline{v_r^2}}{d\log r}$ values negligibly affected on the results.
Even when we substituted $\frac{d\log \overline{v_r^2}}{d\log r}$ with 0 as \citet{2002ApJ...571L..85M} did, the $M^\mathrm{dyn}_\mathrm{250}/M_\mathrm{b}$ values of our ETGs remained larger than those of the spiral galaxies in our sample, and the results did not change.

\begin{figure}
\plottwo{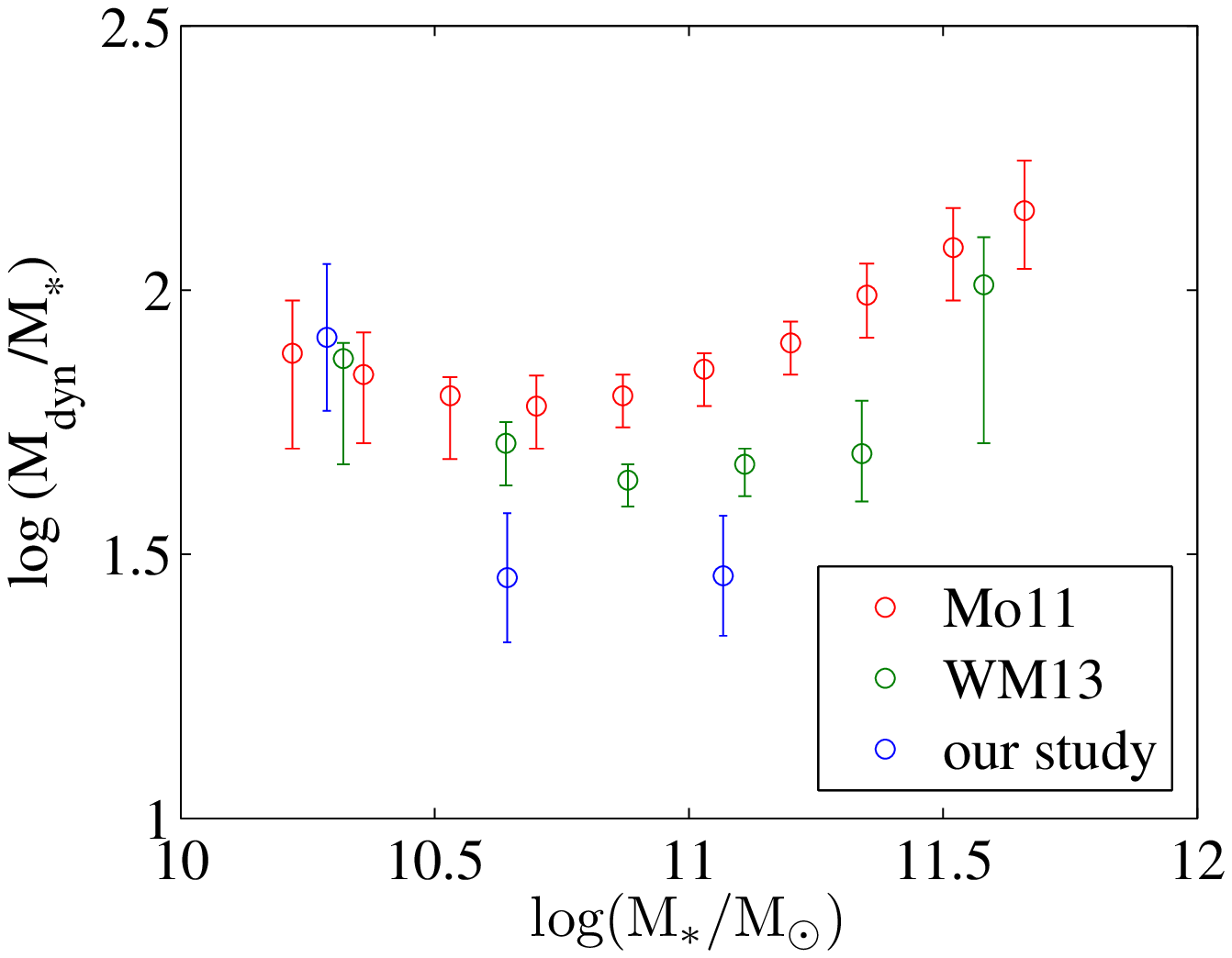}{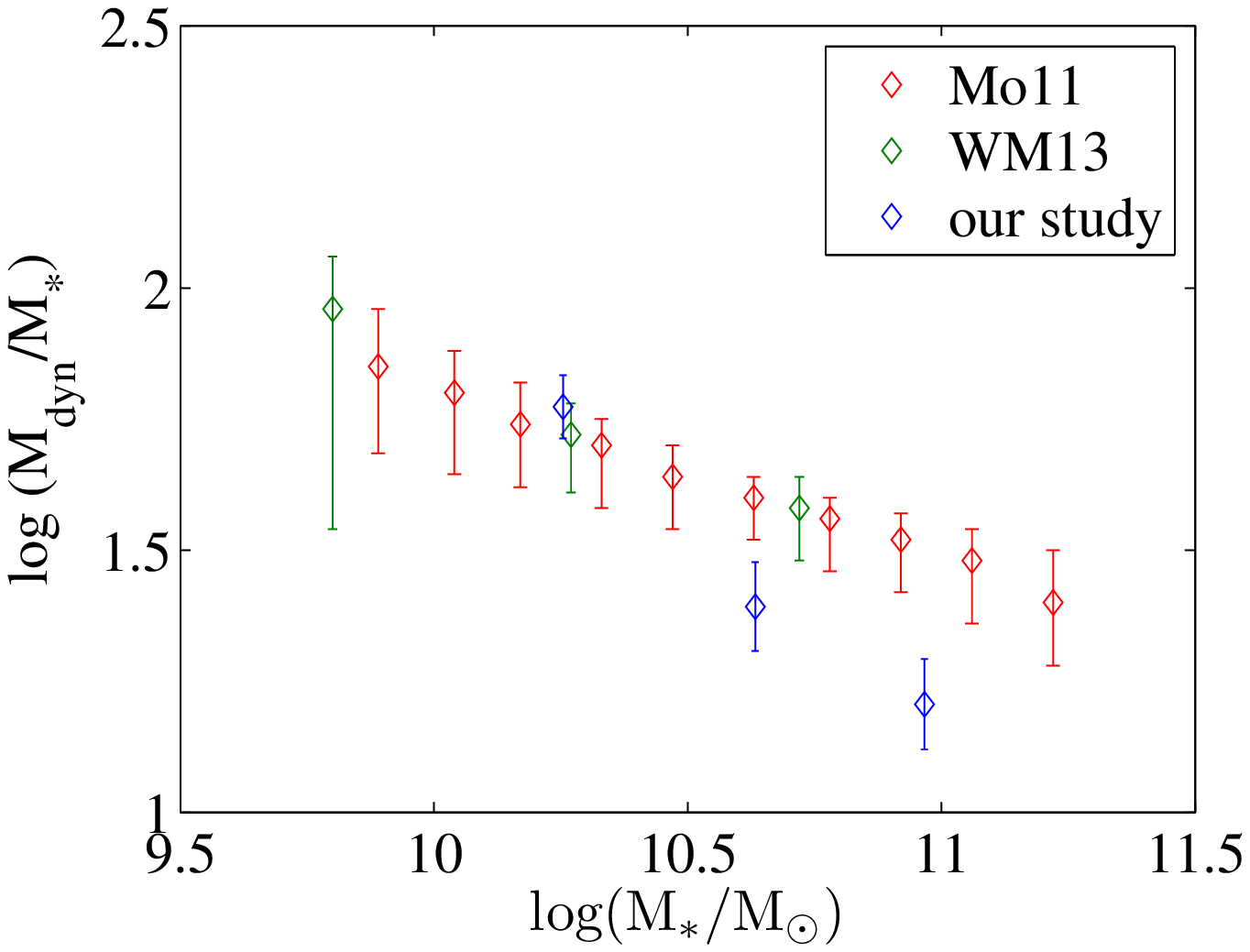}
\caption{
Dynamical-to-stellar mass ratios as a function of stellar mass.
Left: ETGs of our study (blue) and red hosts of \citet[Mo11, red]{2011MNRAS.410..210M} and \citet[WM13, green]{2013MNRAS.428.2407W}.
Right: spirals of our study (blue) and blue hosts of \citet[Mo11, red]{2011MNRAS.410..210M} and \citet[WM13, green]{2013MNRAS.428.2407W}.
$\beta=0$ was assumed for our study in this figure. Please note that the mass in \citet{2011MNRAS.410..210M} and \citet{2013MNRAS.428.2407W} was derived for the mass within 200 times of the critical density.
\label{fig:comparison}}
\end{figure}

Figure \ref{fig:comparison} shows a comparison of our results to the results of \citet{2011MNRAS.410..210M} and \citet{2013MNRAS.428.2407W}. Our results show similar trends to their results but have smaller dynamical-to-stellar mass ratios. The main reason of the difference is likely that we only selected satellite galaxies within $250~\mathrm{kpc}$ of the host galaxies whereas \citet{2011MNRAS.410..210M} and \citet{2013MNRAS.428.2407W} selected satellites within at least $400~\mathrm{kpc}$ of their host galaxies. We note that the mass in \citet{2011MNRAS.410..210M} and \citet{2013MNRAS.428.2407W} was derived for the mass within 200 times of the critical density. The difference of the derived dark matter mass is mainly caused by the different adopted halo sizes.

\begin{figure}
\plotone{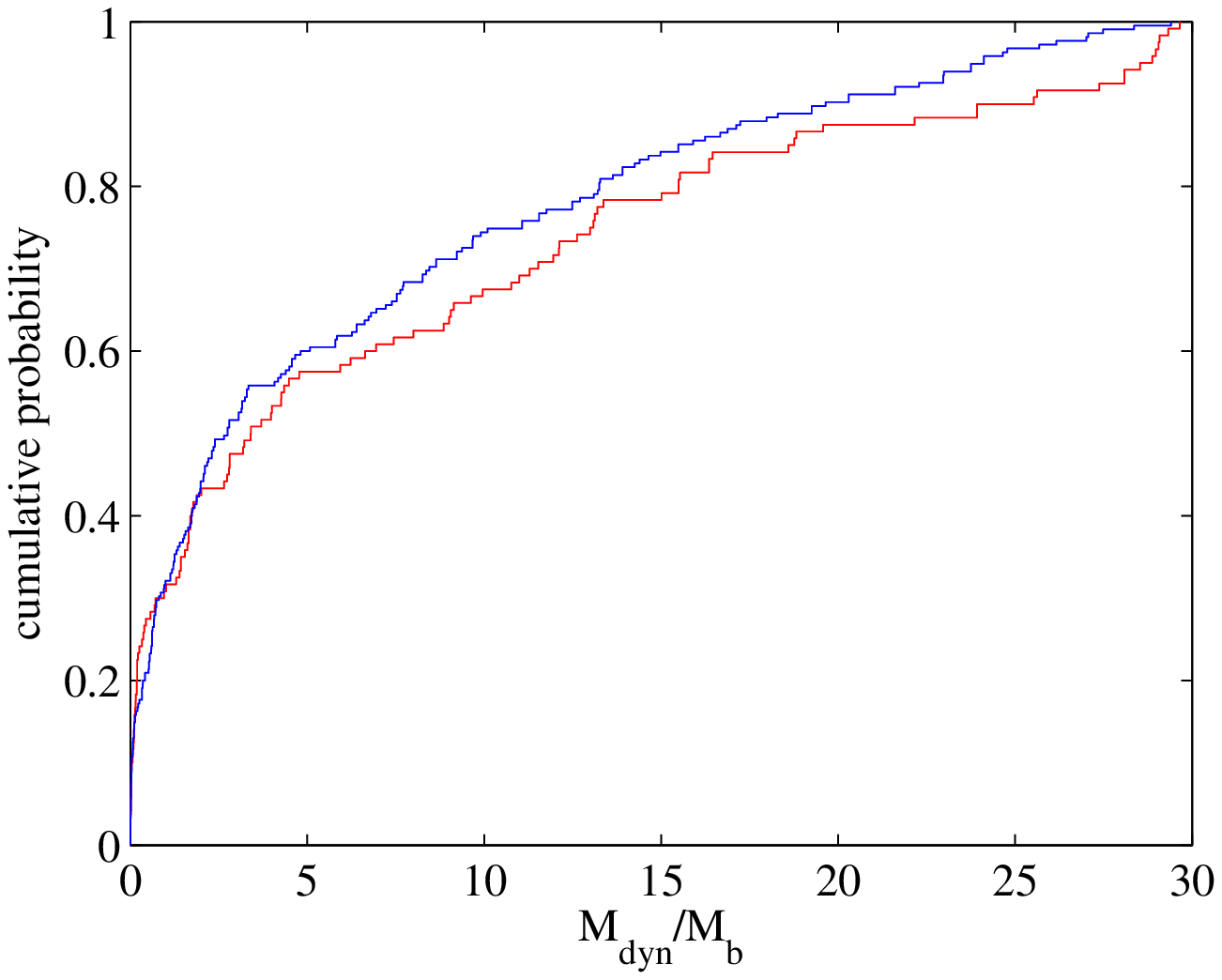}
\caption{Probability distributions of $M_\mathrm{dyn}/M_\mathrm{b}$ of ETGs and spiral galaxies using L17 M--L relation. The red and blue lines represent the distributions of the ETGs and spiral galaxies, respectively.\label{fig:morphology1}}
\end{figure}

\begin{figure}
\plottwo{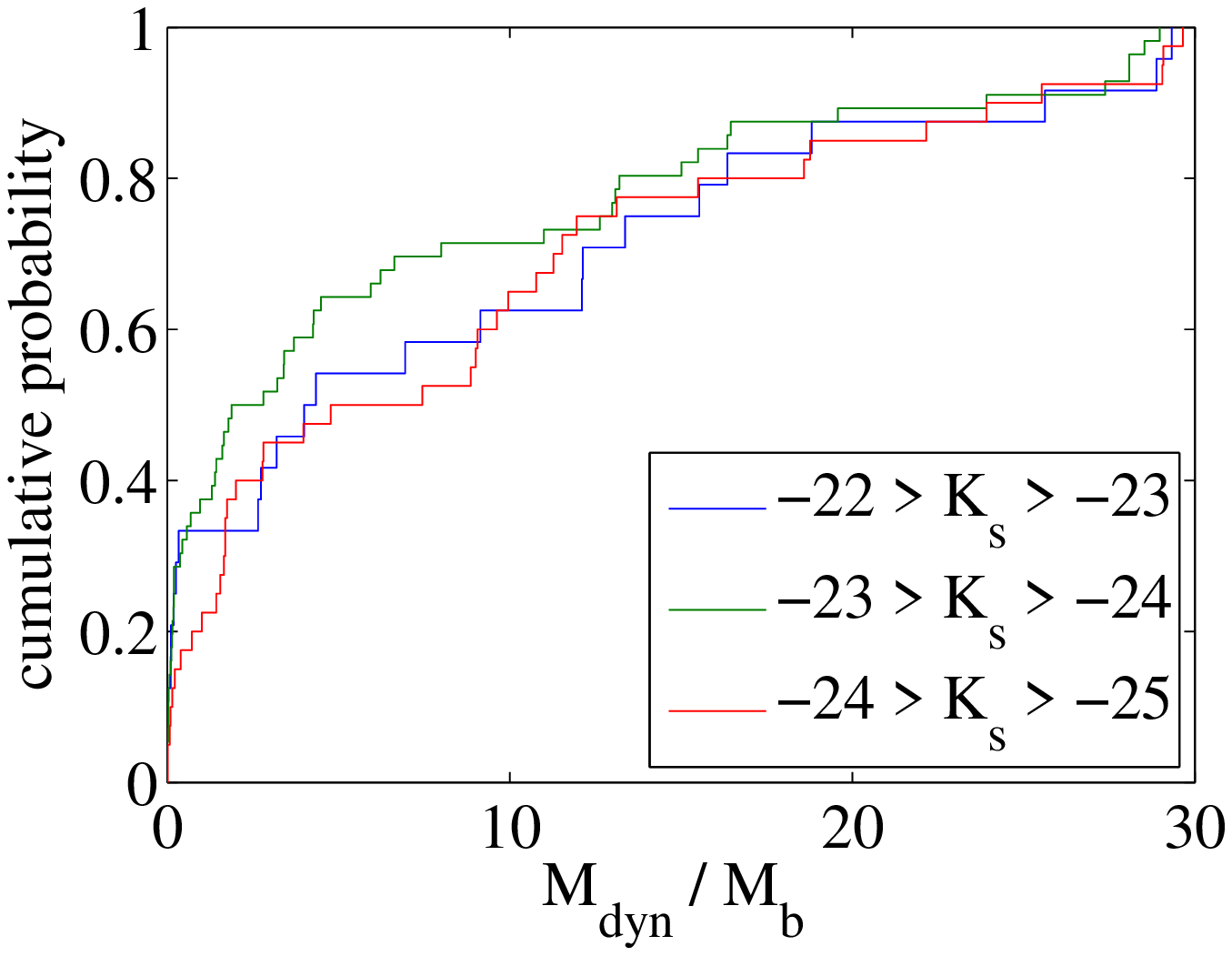}{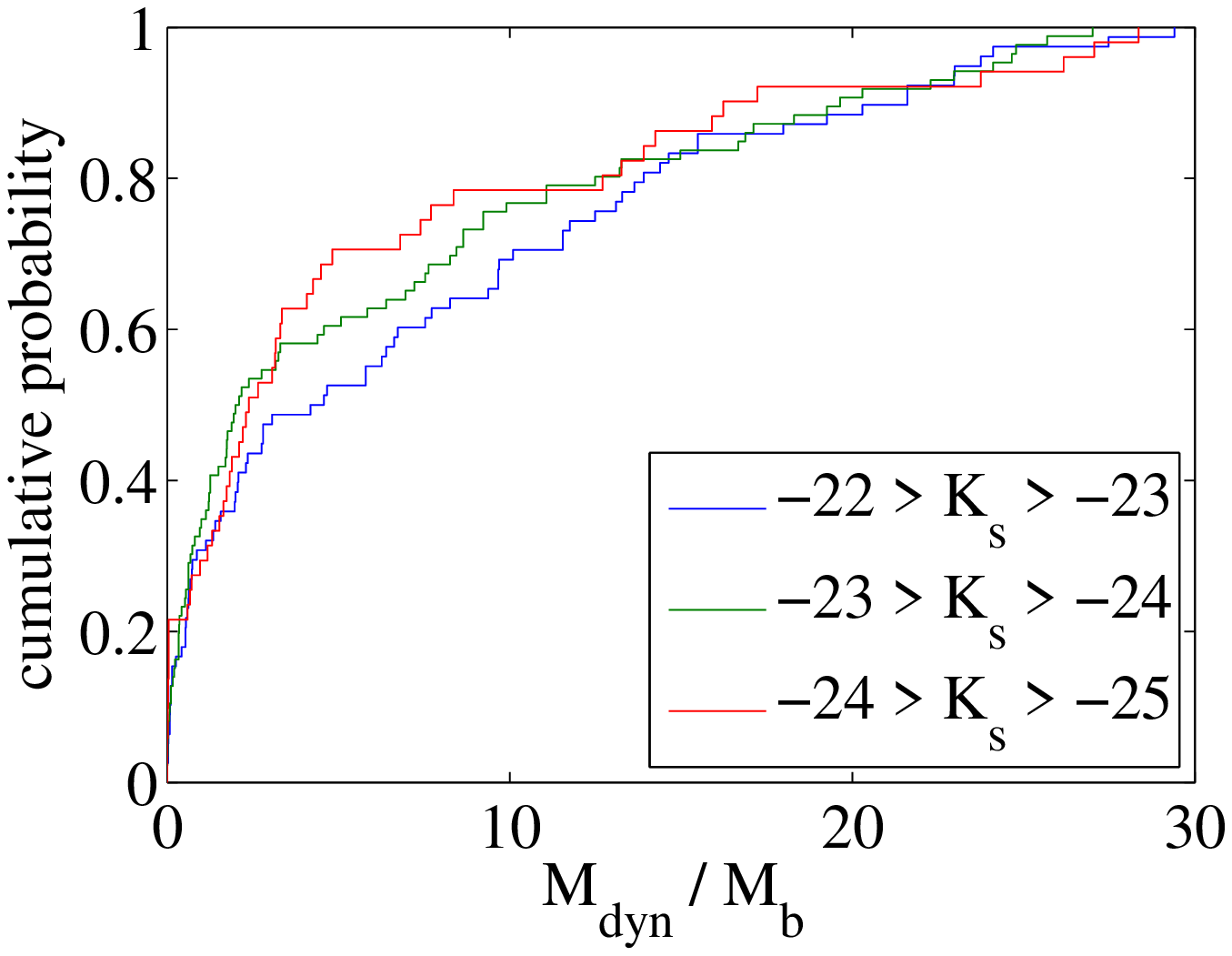}
\caption{Probability distributions of $M_\mathrm{dyn}/M_\mathrm{b}$ of the ETGs and spiral galaxies of different galaxy luminosities using L17 M--L relation.
Left: $M_\mathrm{dyn}/M_\mathrm{b}$ of ETGs.
Right: $M_\mathrm{dyn}/M_\mathrm{b}$ of spiral galaxies.
The blue, green, and red lines represent the $M_\mathrm{dyn}/M_\mathrm{b}$ distributions of the galaxies that had absolute magnitudes from -22 to -23, -23 to -24, and -24 to -25, respectively.\label{fig:luminosity1}}
\end{figure}

\begin{figure}
\plottwo{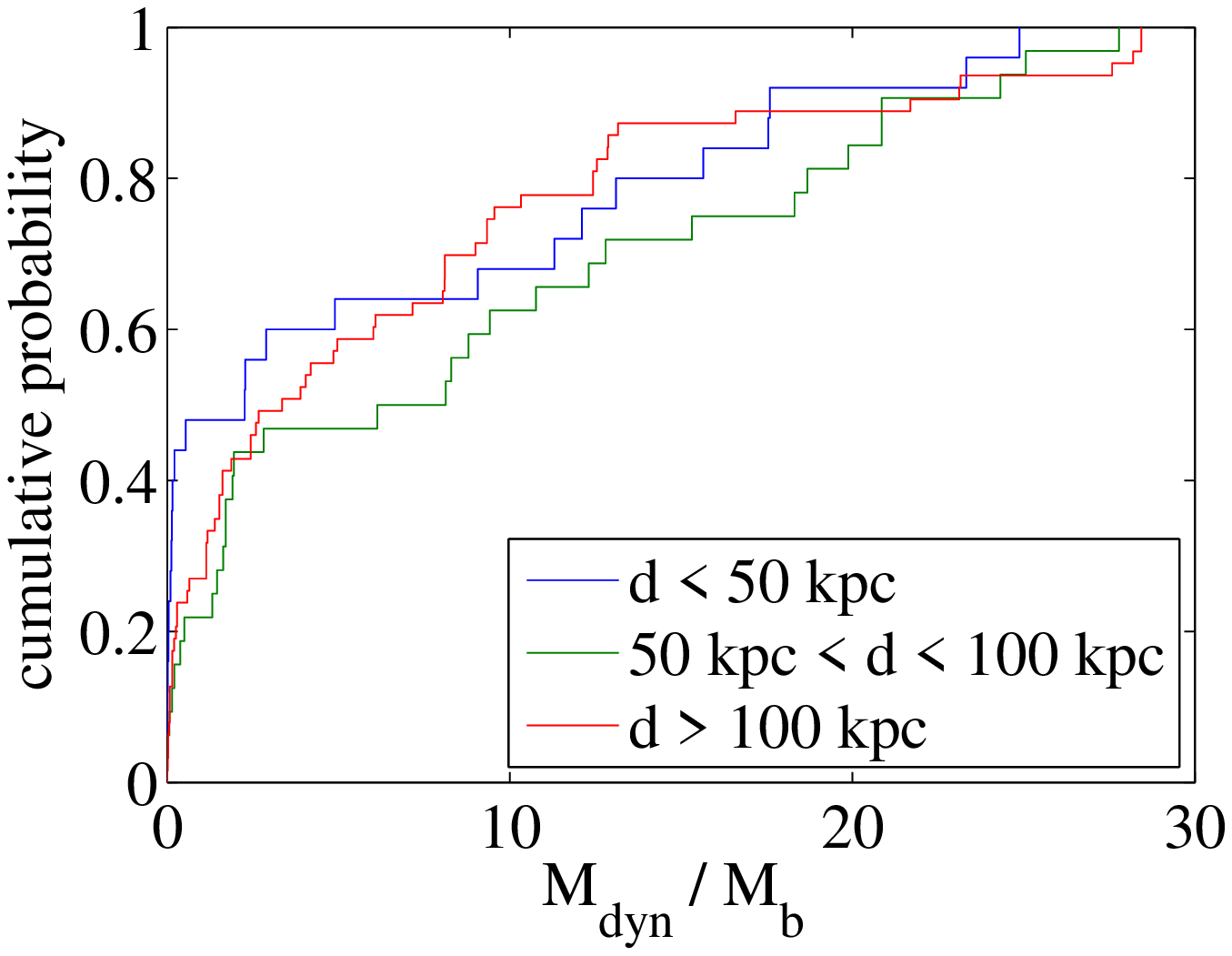}{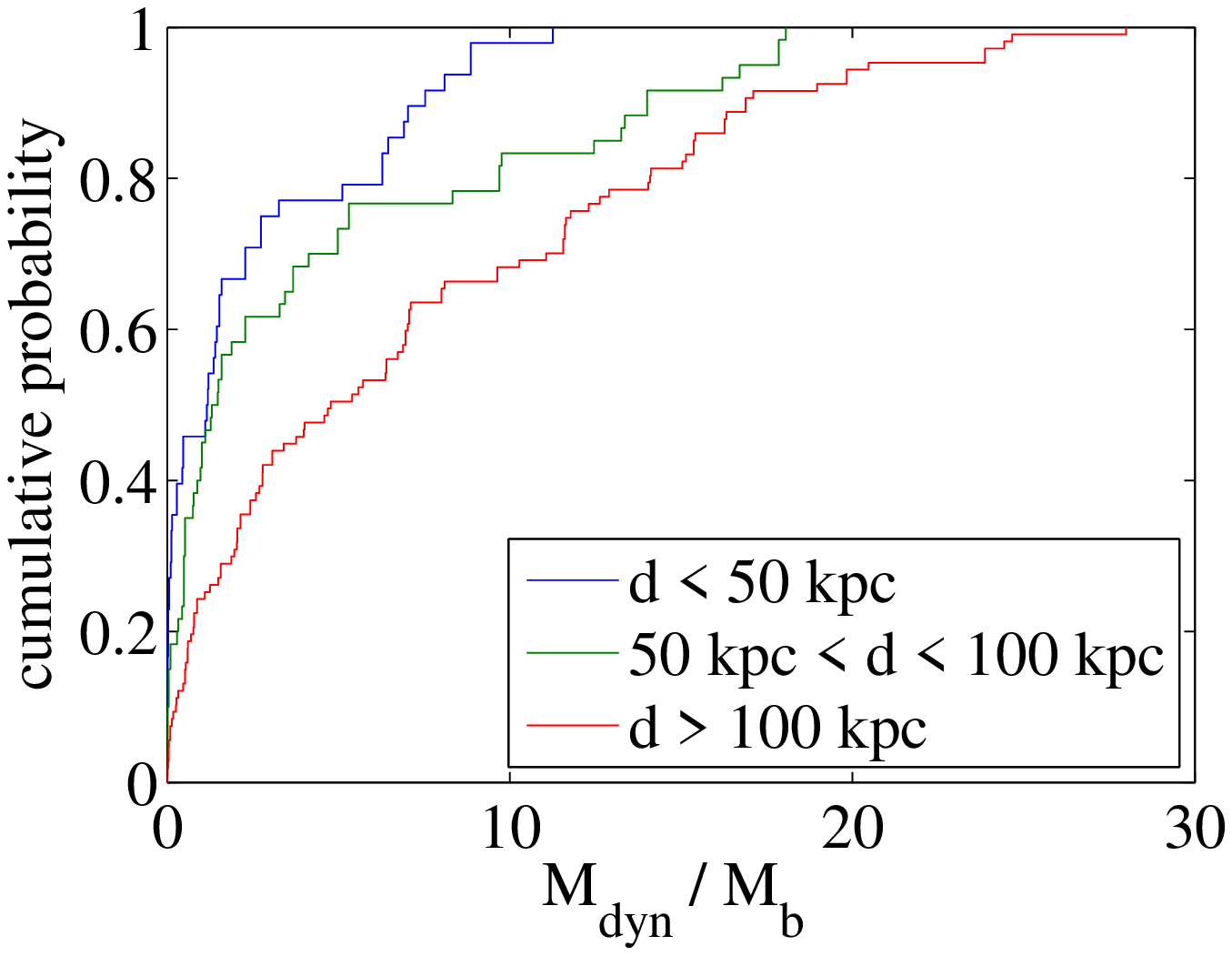}
\caption{Probability distributions of $M_\mathrm{dyn}/M_\mathrm{b}$ of the  ETGs and spiral galaxies of different observed satellite-to-host galaxy distances using L17 M--L relation. Left: $M_\mathrm{dyn}/M_\mathrm{b}$ of ETGs galaxies. Right: $M_\mathrm{dyn}/M_\mathrm{b}$ of spiral galaxies. The blue, green, and red lines represent the $M_\mathrm{dyn}/M_\mathrm{b}$ distributions for distances less than $50~\mathrm{kpc}$, between $50~\mathrm{kpc}$ and $100~\mathrm{kpc}$, and greater than $100~\mathrm{kpc}$, respectively.\label{fig:distance1}}
\end{figure}

We assumed constant stellar $M/L$ ratios to obtain stellar masses.
The luminous mass was dependent on several factors, mainly on the brightness profile and initial mass function (IMF) used to obtain the $M/L$ ratio \citep{2010ApJS..191....1C,2012Natur.484..485C,2013MNRAS.432.1709C,2014MNRAS.445L..79E,2015MNRAS.446...85N,2016MNRAS.462..951N,2019MNRAS.488.1320N}.
We adopted the relationship between stellar mass and luminosity proposed by \citet[hereafter L17]{2017MNRAS.470.2982L} to test the possible effects of these factors.
The L17 stellar mass and luminosity were obtained from the Evolution and Assembly of GaLaxies and their Environments (EAGLE) simulation \citep{2015MNRAS.446..521S,2015MNRAS.450.1937C,2016A&C....15...72M}, which adopts the \citet{2003PASP..115..763C} IMF and the spectral synthesis model of \citet{2003MNRAS.344.1000B} to obtain stellar masses and luminosities, respectively.
The relation is given by
$$\log M = 5.42\times 10^{-2}\exp(0.376\log L)+7.31.$$
This formula was obtained by fitting the published data of L17.
The masses and luminosities are in units of $M_{\sun} h^{-2}$ and $L_{\sun} h^{-2}$, respectively, and we assumed $h=0.7$.
Figure \ref{fig:morphology1} displays the $M_\mathrm{dyn}/M_\mathrm{b}$ distributions of both ETGs and spiral galaxies with the L17 $M/L$ relationship.
The K--S test result revealed a $50\%$ chance of indistinguishable distribution.
We also repeated our analysis for Figures \ref{fig:luminosity} and \ref{fig:distance} using the L17 $M/L$ relationship; the results are illustrated in Figures \ref{fig:luminosity1} and \ref{fig:distance1}.
The K--S test of the distributions is displayed in Table \ref{tab:KStest1}, which indicate that the results illustrated in Figures \ref{fig:luminosity1} and \ref{fig:distance1} are similar to those displayed in Figure \ref{fig:luminosity} and \ref{fig:distance}.
These findings suggest that the similarity in $M_\mathrm{dyn}/M_\mathrm{b}$ ratios between ETGs and spiral galaxies were not affected by differences in luminous mass estimates.

In this study, we compared the dynamics of the companion galaxies of ETGs and spiral galaxies.
Our results revealed that first, ETGs have higher dynamical-to-baryonic mass ratios than spiral galaxies have at the same baryonic mass.
Second, satellites of ETGs tend to follow more elliptical orbits than satellites of spiral galaxies do.
We noted that these conclusions were derived statistically and may not apply to individual galaxies.

\acknowledgments

This work is supported by the Ministry of Science and Technology of Taiwan (grants MOST 107-2119-M-008-009-MY3).
This research has made use of the NASA/IPAC Extragalactic Database (NED), which is operated by the Jet Propulsion Laboratory, California Institute of Technology, under contract with the National Aeronautics and Space Administration.

\clearpage

\newpage

\clearpage

\newpage

\begin{table}[htbp]
  \centering
  \caption{Probabilities of the K--S test of $M_\mathrm{dyn}/M_\mathrm{b}$ in various observed satellite-to-host galaxy distances.}
    \begin{tabular}{c|cc|cc}
    \hline
    \hline
    \multicolumn{1}{r}{} & \multicolumn{2}{c}{ETGs} & \multicolumn{2}{c}{spiral} \\
          & $d<50~\mathrm{kpc}$  & $50~\mathrm{kpc}<d<100~\mathrm{kpc}$ & $d<50~\mathrm{kpc}$  & $50~\mathrm{kpc}<d<100~\mathrm{kpc}$ \\
    \hline
    $50~\mathrm{kpc}<d<100~\mathrm{kpc}$ & $20\%$ & -- & $4.6\%$ & -- \\
    $d>100~\mathrm{kpc}$ & $35\%$ & $64\%$ & $0.19\%$ & $2.4\%$ \\
    \hline
    \end{tabular}
  \label{tab:KStest}
\end{table}

\clearpage

\newpage

\clearpage

\newpage

\clearpage

\newpage

\clearpage

\newpage

\clearpage

\newpage

\begin{table}[htbp]
  \centering
  \caption{Results of the K--S test of $M_\mathrm{dyn}/M_\mathrm{b}$ in various observed satellite-to-host galaxy distances.}
    \begin{tabular}{c|cc|cc}
    \hline
    \hline
    \multicolumn{1}{r}{} & \multicolumn{2}{c}{ETGs} & \multicolumn{2}{c}{spiral} \\
          & $d<50~\mathrm{kpc}$  & $50~\mathrm{kpc}<d<100~\mathrm{kpc}$ & $d<50~\mathrm{kpc}$  & $50~\mathrm{kpc}<d<100~\mathrm{kpc}$ \\
    \hline
    $50~\mathrm{kpc}<d<100~\mathrm{kpc}$ & $20\%$ & -- & $31\%$ & -- \\
    $d>100~\mathrm{kpc}$ & $39\%$ & $36\%$ & $0.18\%$ & $2.8\%$ \\
    \hline
    \end{tabular}
  \label{tab:KStest1}
\end{table}

\clearpage

\appendix

\section{Simulation for observed mass ratio distribution}
\begin{figure}
\plotone{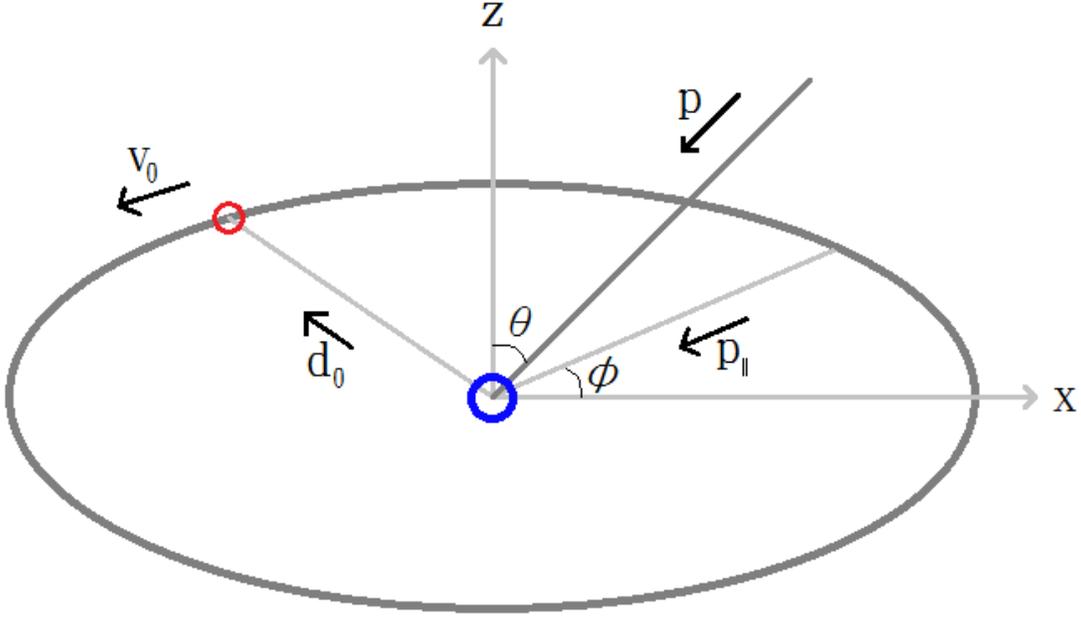}
\caption{
Schematic of the line-of-sight and the host-satellite plane.
The blue and red circles represent the host and satellite galaxy, repectively.
The z-axis is the rotation axis of the satellite galaxy.
The direction of the real velocity is $\vec{v_0}$, the real distance is $\vec{d_0}$, the line-of-sight is $\vec{p}$, and the projecting component of $\vec{p}$ on host-satellite plane is $\vec{p_{\parallel}}$.
${\phi}$ is the angle between x-axis and $\vec{p_{\parallel}}$;
${\theta}$ is the angle between z-axis and $\vec{p}$.
\label{fig:orbit}}
\end{figure}

For a host galaxy with mass $M_0$, a satellite galaxy at a circular orbit with distance $d_0$ to the host galaxy will have a rotation velocity $v_0$: 
\begin{equation}
v_0^2 = \frac{GM_0}{d_0}.
\end{equation}
For an observer in a random position, the unit vector of the line-of-sight $\vec{p}$ is related to the origin of the host galaxy with
the following equations
\begin{eqnarray}
p_x & = & \cos\phi\sqrt{1-p_z^2}, \nonumber \\
p_y & = & \sin\phi\sqrt{1-p_z^2}, \nonumber \\
p_z & = & 2\mu - 1,
\end{eqnarray}
where $\mu=\cos\theta$, $\phi$ and $\theta$ are the parameters of $\vec{p}$ in the polar coordinate centred at the origin of the host galaxy.
Figure \ref{fig:orbit} displayed a schematic of the line-of-sight and the host-satellite plane.
The observed separate distance $d_1$ is the projected component perpendicular to the line-of-sight, and the observed velocity $v_1$ is the projected velocity along the line-of-sight.
Assuming that the observed velocity square represents one-third of the true velocity square,
the observed mass of the host galaxy $M_1$ has the following relationship to the true mass $M_0$
\begin{equation}
M_1 = \frac{3v_1^2 d_1}{G} = M_0 \frac{3v_1^2}{v_0^2} \frac{d_1}{d_0}.
\label{M1}
\end{equation}
Therefore the maximum $M_1$ could be three $M_0$ in the case of the line-of-sight is parallel to the direction of $\vec{v_0}$.
For a host galaxy at the original point and a satellite orbit on the $x-y$ plane, $d_1$ and $v_1$ are functions of $\vec{d}$, $\vec{v}$ and $\vec{p}$.
\begin{eqnarray}
d_1 & = & \sqrt{d_0^2-(\vec{d_0}\cdot\vec{p})^2}, \nonumber \\
v_1 & = & |\vec{v_0}\cdot\vec{p}|.
\label{d1v1}
\end{eqnarray}

If we put a host galaxy with mass $M_0$ at the original point and a satellite at $(d_0,0,0)$ with initial velocity $(0,v,0)$ where $v = k v_0$, we get a circular orbit for $k=1$ and elliptical orbits for $0<k<\sqrt{2}$ but $k \neq 1$.
Moreover, the initial point would be the apocenter for $0<k<1$ and pericenter for $1<k\sqrt{2}$.

With the same ellipticities, we can have different $d_0$ and $v_0$.
We scaled real distance and velocity of the satellite to $cd_0$ and $v/\sqrt{c}$, respectively.
The range of $c$ in our simulation is showed in Table \ref{tab:parameter}
We note that we have assumed the mass distribution of the host galaxy to be with the pericenter distance of the minima orbit of the satellite in this simulation.

In our study, we used five different $k$, which were all less than or equal to one. Therefore, the initial points are the apocenter points.
We first used $M_0=10^{11}M_{\sun}$ (assuming $M_*=10^{10}M_{\sun}$) and $d=250~\mathrm{kpc}$ to simulate the positions and velocities of the satellite galaxy for the five orbits.
For each orbit, we then randomly selected a position, a sight vector $\vec{p}$ and a scaled parameter $c$ to simulate an observation event.
The parameters $\mu$ and $\phi$ are assumed to be uniformly distributed in $[0,1]$ and $[0,2\pi]$, respectively.
The observed $d_1$ and $v_1$ can be determined by the Equation \ref{M1}, and the observed $M_1$ can be determined by the Equation \ref{d1v1}.
We simulated one million observation events for each orbit; the results are illustrated in Figure \ref{fig:sim}.
Table \ref{tab:parameter} displayed the parameters for the five orbits which we used in this study.

\begin{table}[htbp]
  \centering
  \caption{Parameters of the simulated orbits}
    \begin{tabular}{cccccc}
    \hline
    \hline
    $k$ & pericenter distance & $\epsilon$ & range of c \\
    & (kpc) & & \\
    \hline
    1 & 250 & 1 & $[1/5,1]$ \\
    0.943 & 200 & 0.8 & $[1/4,1]$ \\
    0.866 & 150 & 0.6 & $[1/3,1]$ \\
    0.756 & 100 & 0.4 & $[1/2,1]$ \\
    0.577 & 50 & 0.2 & $[1,1]$ \\
    \hline
    \end{tabular}
  \label{tab:parameter}
\end{table}

\end{document}